\documentclass[11pt,letterpaper]{article}
\pdfoutput=1

\usepackage{jcappub}

\usepackage{color}
\usepackage{graphicx}
\usepackage[space]{grffile}
\usepackage{ulem}

\usepackage{verbatim}
\usepackage{amsmath}
\usepackage{amssymb}
\usepackage[caption=false]{subfig}
\usepackage{url}
\usepackage{bbold}
\usepackage{slashed}
\usepackage{cancel}
\usepackage{epstopdf}
\usepackage{wasysym}
\usepackage{soul}

\usepackage{multirow}
\usepackage{threeparttable}
\usepackage{paralist}
\usepackage{tikz}

\newcommand{\MeV}{\text{MeV}}
\newcommand{\GeV}{\text{GeV}}
\newcommand{\TeV}{\text{TeV}}

\newcommand{\pythia}{{\sc Pythia}\xspace}

\DeclareRobustCommand{\Sec}[1]{Sec.~\ref{#1}}

\DeclareRobustCommand{\App}[1]{App.~\ref{#1}}
\DeclareRobustCommand{\Tab}[1]{Table~\ref{#1}}

\DeclareRobustCommand{\Fig}[1]{Fig.~\ref{#1}}
\DeclareRobustCommand{\Figs}[2]{Figs.~\ref{#1} and \ref{#2}}
\DeclareRobustCommand{\Eq}[1]{Eq.~(\ref{#1})}
\DeclareRobustCommand{\Eqs}[2]{Eqs.~(\ref{#1}) and (\ref{#2})}
\DeclareRobustCommand{\Ref}[1]{Ref.~\cite{#1}}
\DeclareRobustCommand{\Refs}[1]{Refs.~\cite{#1}}

\newcommand{\be}{\begin{equation}}
\newcommand{\ee}{\end{equation}}
\newcommand{\bea}{\begin{eqnarray}}
\newcommand{\eea}{\end{eqnarray}}
\newcommand{\bi}{\begin{itemize}}
\newcommand{\ei}{\end{itemize}}

\newcommand{\ra}{\rightarrow}
\newcommand{\ov}{\overline}
\newcommand{\eq}{{\rm{eq}}}
\newcommand{\da}{\dagger}

\usepackage{xspace}

\newcommand{\Abar}{\overline{A}}

\newcommand{\Bbar}{\overline{B}}

\newcommand{\lp}{\left(}
\newcommand{\rp}{\right)}

\newcommand{\lang}{\langle}
\newcommand{\rang}{\rangle}

\newcommand{\pbar}{\overline{p}}

\newcommand{\df}{\mathrm{d}}

\makeindex

\begin{document}

\title{Dark Matter, Shared Asymmetries, and  Galactic Gamma Ray Signals}

\author[a,b]{Nayara Fonseca,}

\author[b]{Lina Necib,}

\author[b]{and Jesse Thaler}

\affiliation[a]{
Instituto de F\'{i}sica, Universidade de S\~{a}o Paulo, S\~{a}o Paulo, SP 05508-900, Brazil }

\affiliation[b]{Center for Theoretical Physics, Massachusetts Institute of Technology, Cambridge, MA 02139, USA}

\emailAdd{nayara@if.usp.br}
\emailAdd{lnecib@mit.edu}
\emailAdd{jthaler@mit.edu}

\date{today}

\abstract{We introduce a novel dark matter scenario where the visible sector and the dark sector share a common asymmetry.  The two sectors are connected through an unstable mediator with baryon number one, allowing the standard model baryon asymmetry to be shared with dark matter via semi-annihilation.  The present-day abundance of dark matter is then set by thermal freeze-out of this semi-annihilation process, yielding an asymmetric version of the WIMP miracle as well as promising signals for indirect detection experiments.  As a proof of concept, we find a viable region of parameter space consistent with the observed \textit{Fermi} excess of GeV gamma rays from the galactic center.}

\preprint{MIT-CTP {4695}}

\maketitle

\section{Introduction}
\label{sec:intro}

The existence of dark matter (DM) has been confirmed through many independent studies, and we now know that this non-baryonic matter comprises around a fifth of the energy density of the universe  \cite{Zwicky:1933gu,Bertone:2010zza,Planck:2015xua}.   If DM is composed of a single new particle, then the Weakly Interacting Massive Particle (WIMP) paradigm is particularly attractive, since the present-day abundance of DM can be set by thermal freeze-out in an expanding universe (see \cite{Bertone:2004pz} for a review).  Given that the standard model (SM) features a vast and subtle structure, though, there may be comparably rich dynamics in the dark sector.  Perhaps the dark sector features some of the same properties as visible SM baryons, such as accidental symmetries, asymmetric abundances, and instability of some of its components.

In this paper, we propose a new DM scenario where asymmetries in the visible and dark sectors are closely connected.  Due to light unstable states carrying baryon number, asymmetry sharing is efficient down to temperatures below the DM mass, such that the ultimate (asymmetric) abundance of DM is set by thermal freeze-out of a semi-annihilation process. This can be regarded as a hybrid framework between asymmetric and WIMP  scenarios, whereby the processes responsible for asymmetry sharing in the early universe  can potentially produce signals today in indirect detection experiments.

As a prototypical example, consider a stable DM particle $A$ which carries baryon number $1/2$.  This DM particle couples to an unstable particle $B$ with baryon number $1$ and to a light mediator $\phi$ with baryon number $0$.  The semi-annihilation \cite{Hambye:2008bq, Hambye:2009fg, Arina:2009uq, DEramo:2010ep, Belanger:2012vp, Aoki:2012ub}  process
\be
\label{eq:semiannihilation}
AA \to B \phi
\ee
allows the SM baryon asymmetry to be shared in the early universe, and it also gives rise to promising indirect detection signals in the galactic halo today.  Note that this process does not involve $\Abar$, since this is an asymmetric DM scenario.  As we will see, there is a viable region of parameter space that not only yields the desired DM and baryon abundances but also yields a GeV gamma ray signal consistent with the galactic center (GC) excess seen in \textit{Fermi} \cite{Goodenough:2009gk, Hooper:2010mq, Hooper:2011ti, Abazajian:2012pn, Hooper:2013rwa, Gordon:2013vta, Huang:2013pda, Macias:2013vya, Abazajian:2014fta,  Daylan:2014rsa, Zhou:2014lva, Calore:2014xka}.

\begin{figure}[t]
   \centering
   \subfloat[]{{\includegraphics[scale=0.6]{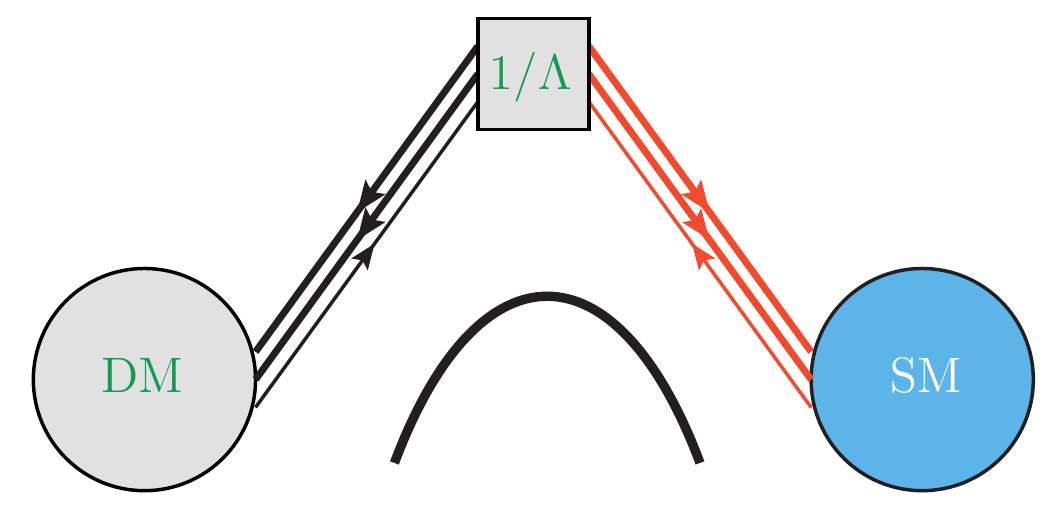}  } \label{fig:intro:a} }
   \qquad \qquad
   \subfloat[]{{\includegraphics[scale=0.6]{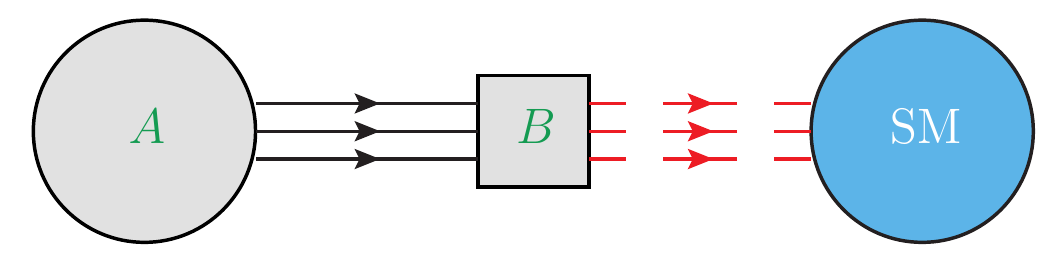} }  \label{fig:intro:b} }
   \caption{ Two approaches to sharing an asymmetry between the DM and SM sectors: (a) through a higher-dimension operator that becomes inefficient as the universe cools, or (b) though an on-shell intermediate particle $B$ which remains in quasi-equilibrium until $A$ freezes out.  In this paper, we assume that $B$ chemically decouples from the SM before $A$ freezes out (indicated by the dashing), in which case freeze-out of semi-annihilation $AA \rightarrow B \phi$ controls the final DM abundance.}
   \label{fig:intro}
\end{figure}

The presence of asymmetry sharing is familiar from asymmetric DM models (see \cite{Petraki:2013wwa,Zurek:2013wia} for reviews), but there is a key difference highlighted in \Fig{fig:intro}.\footnote{We assume that the initial baryon asymmetry is determined by some unspecified high scale process, in either the visible sector, the baryon sector, or both.   There are also interesting scenarios where the baryon asymmetry is generated via the freeze-out process itself \cite{Cui:2011ab,  Bernal:2012gv, Cui:2012jh, Bernal:2013bga, Racker:2013lua}, but we will not consider that here.}   In scenarios shown in \Fig{fig:intro:a}, some high-dimension operator or heavy state links the visible and dark sectors \cite{Kaplan:2009ag} (see also \cite{Cohen:2009fz, Cohen:2010kn,Shelton:2010ta, Davoudiasl:2010am, Haba:2010bm,Buckley:2010ui,  McDonald:2011zza, Arina:2011cu,McDonald:2011sv, Servant:2013uwa}).  As the temperature of the universe drops, this interaction becomes irrelevant, locking in independent asymmetries for baryons and DM.  In our scenario shown in \Fig{fig:intro:b}, the interactions that link the visible and dark sectors involve light mediators in quasi-equilibrium, and therefore stay relevant to energies below the DM mass.  As long as $m_B < 2 m_A$, \Eq{eq:semiannihilation} is always kinematically allowed, impacting both thermal freeze-out in the early universe and indirect detection today.

There are three relevant temperatures in this scenario.  The decoupling temperature $T_D$ is when the mediator $B$ chemically decouples from SM baryons, setting the asymmetries in the dark sector.  The effective symmetry imbalance temperature $T_I$ is when the abundance of $\Abar$ (and $\Bbar$) departs significantly from the abundance of $A$ (and $B$).  The freeze-out temperature $T_F$ is when the semi-annihilation process $AA \rightarrow B \phi$ freezes out, setting the final $A$ abundance.  We will focus on the hierarchy
\be
T_D > T_I \gtrsim T_F,
\ee
which yields an asymmetric version of the WIMP miracle.  As we will see, achieving $T_I \gtrsim T_F$ requires 
\be
\label{eq:massrelation}
m_A \lesssim m_B < 2 m_A.
\ee
With this mass range it is natural (though not required) to consider $B$ as a bound state of two $A$ particles.  For example, the semi-annihilation process in \Eq{eq:semiannihilation} could correspond to the formation of dark bound states \cite{Pearce:2013ola}, dark nuclei \cite{Krnjaic:2014xza,Detmold:2014qqa}, or dark atoms \cite{Pearce:2015zca}, though the key difference compared to previous work is that the resulting $B$ particle is unstable.\footnote{To achieve a more predictive framework, one could consider $B$ to be a bound state of $A$ particles mediated by $\phi$ exchange.  In this case, $m_B$ would be determined by $m_A$, $m_\phi$, and the $A$-$\phi$ coupling, and there would be tight relationship between the various interaction cross sections.  In order to explore the full parameter space, we will not assume that $B$ is a bound state resulting from $\phi$ interactions.  Indeed, in dark nuclei scenarios \cite{Krnjaic:2014xza,Detmold:2014qqa}, $B$ is bound because of (residual) confining interactions and $\phi$ is a spectator to the binding process.}

Unlike standard asymmetric DM models, \Eq{eq:semiannihilation} leads to indirect detection signals without requiring a residual symmetric component.  The resulting indirect detection spectrum depends on the decays of $B$ (and to a lesser extent, the decays of $\phi$).  As benchmarks, we consider two possibilities involving SM quarks,\footnote{These kinds of decay modes also appeared in the asymmetric DM scenario of \Ref{Zhao:2014nsa}.   In that model, $B$ itself was DM and had late-time decays.  Here, $A$ is DM and $B$ decays are prompt on cosmological scales.}
\be
\label{eq:decaymodes}
B \to u d d, \qquad B \to c b b,
\ee
and both of these decay modes can yield a good fit to the \textit{Fermi} GC excess, albeit with different preferred values of $m_A$ and $m_B$.  As we will see, the desired phenomenology requires $m_A$ and $m_B$ to have comparable masses of $\mathcal{O}(10\text{--}100~\GeV)$.\footnote{The astute reader may notice that the above processes do not conserve fermion number.  This is easily reconciled with an extended model presented in \App{app:model}, yielding equivalent phenomenology.  Similarly, $\phi$ can correspond to multiple unstable states, specifically two in \App{app:model}.}

\begin{figure}[t]
   \centering
   \subfloat{
   \includegraphics[height=9cm,keepaspectratio]{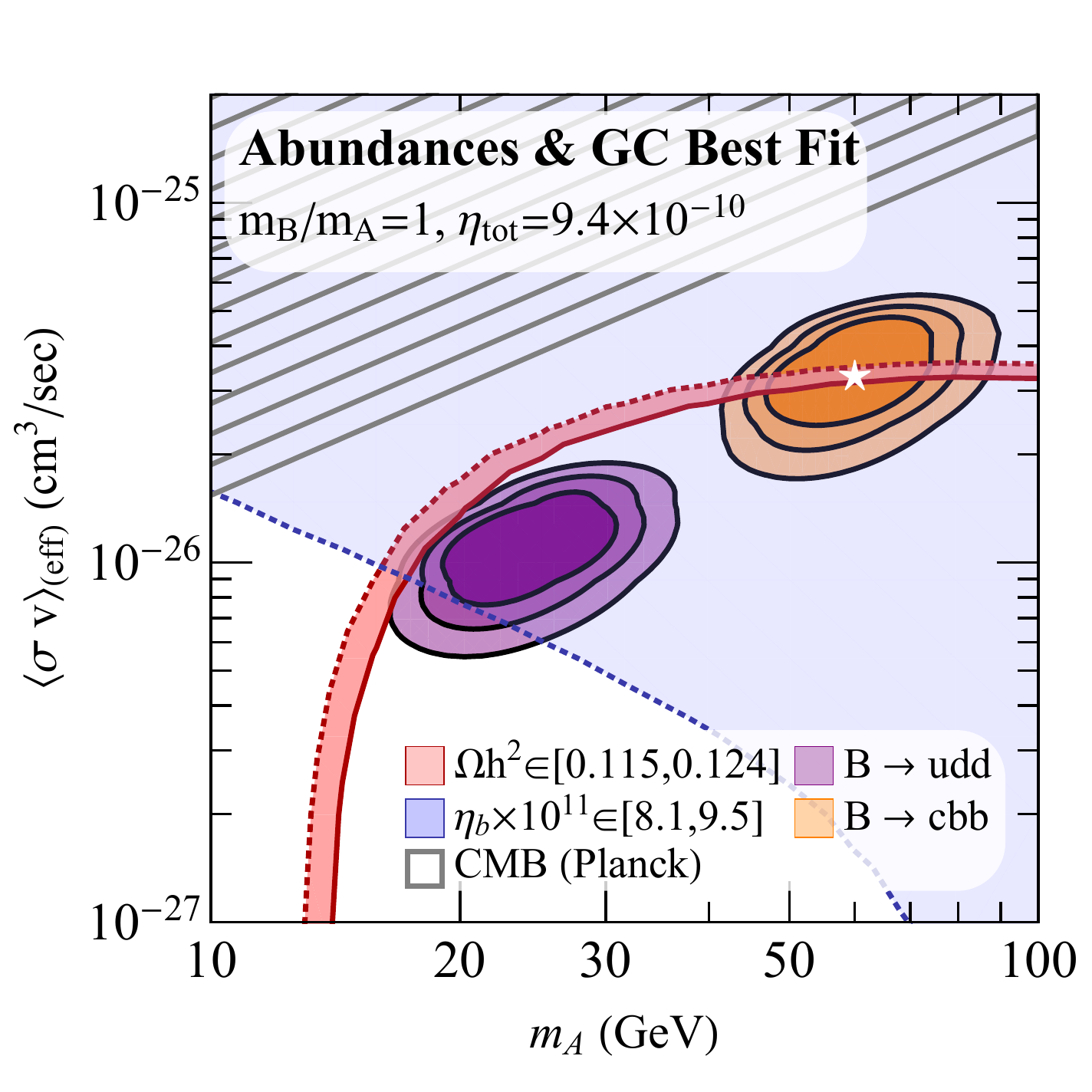} 
   }
   \caption{Example region of parameter space where we obtain the desired DM abundance \cite{Planck:2015xua} (red), the desired baryon asymmetry \cite{Agashe:2014kda} (blue), and a good fit to the GC excess from $AA \to B \phi$ semi-annihilation followed by $B \to udd$ (purple) or $B \to cbb$ (orange) decays.  Here, $\Omega_{\rm DM} h^2$ and $\eta_b$ are plotted with respect to the true cross section $\langle \sigma_{AA \to B \phi} v \rangle$, while the GC fit regions are shown with respect to the effective cross section $\langle \sigma_{AA \to B \phi} v \rangle_\text{eff}$ scaled by the DM abundance (see \Eq{eq:sigmaeff}).   The hashed region indicates the CMB heating bounds from \Eq{eq:CMB}, evaluated with respect to $\langle \sigma v \rangle_\text{eff}$.  Not shown are the AMS-02 antiproton bounds from \Eq{eq:antiprotonbounds}, which do constrain this parameter space but have large uncertainties.  The white star indicates the benchmark parameters in \Eq{eq:benchmark}.}
   \label{fig:moneyplotintro}
\end{figure}  

As a proof of concept, we highlight a benchmark scenario that satisfies the following three criteria, summarized in \Fig{fig:moneyplotintro}.
\begin{itemize}
\item \textit{Correct DM abundance}: We require $\Omega_{\text{DM}}h^2 \in [0.115, 0.124]$, within the  95$\%$ Planck confidence
 limit of the DM abundance  \cite{Planck:2015xua}.\footnote{For simplicity, we assume a Gaussian distribution to extrapolate  the  $2\, \sigma$ Planck range from the $1\, \sigma$  limit ($ \Omega_{\text{DM}}h^2 \in[0.1175, 0.1219]$)  in  \Ref{Planck:2015xua}.  } 
\item \textit{Correct baryon asymmetry}: We require $\eta_b = (n_b - n_{\overline{b}})/s \in [8.1, 9.5] \times 10^{-11}$, within the 95$\%$ confidence limit of the baryon asymmetry of the universe \cite{Agashe:2014kda}.\footnote{Here, $s$ is the entropy density.  This differs from the more familiar notation for $\eta_b$, which is defined with respect to the photon density. Today $s = 7.05 n_\gamma$ \cite{Agashe:2014kda,Kolb:1990vq}.  }
\item \textit{Plausible fit to the GC excess}:  We follow the strategy of \Ref{Calore:2014xka} to find the best fit masses for the GC excess gamma ray spectrum. 
\end{itemize}
Simultaneously satisfying these criteria turns out to be possible, since much of the intuition derived from symmetric WIMPs holds also in this asymmetric scenario.  We check that this benchmark is consistent with CMB heating bounds from \textit{Planck} \cite{Planck:2015xua}, and we evaluate possible direct detection and antiproton flux constraints.  We also investigate searches for displaced jets at the Large Hadron Collider (LHC), which could be a distinctive signature for the proposed scenario in optimistic regions of parameter space. 

The rest of this paper is organized as follows.  We introduce our prototype model in \Sec{sec:model} and study its cosmological history and asymmetry sharing in \Sec{sec:earlyhistory}, leaving details to the appendices.   In \Sec{sec:GC}, we show how the $AA \to B \phi$ semi-annihilation process followed by $B$ decay can match the observed GC excess gamma ray spectrum.   We discuss constraints and potential collider signals in \Sec{sec:constraints} and conclude in \Sec{sec:conclude}.

\section{Prototype of Dark Matter with a Shared Asymmetry}
\label{sec:model}

The analysis in this paper is based on a simplified dark sector consisting of three particles:  a stable DM species $A$ with baryon number $1/2$, an unstable state $B$ with baryon number $1$, and a light mediator $\phi$ which couples to the SM through, e.g., the hypercharge portal \cite{Holdom:1985ag,Okun:1982xi,Galison:1983pa,Pospelov:2007mp,ArkaniHamed:2008qn}, Higgs portal \cite{Schabinger:2005ei,Patt:2006fw}, or axion portal \cite{Nomura:2008ru,Mardon:2009gw}.  The reason the $A$ particle is exactly stable is that there are no states in the SM with baryon number $1/2$.  An explicit Lagrangian with these properties is presented in \App{app:model}, where the state $A$ is replaced with a fermion/boson system.  For simplicity, we take this $\psi_A$/$\phi_A$ system be mass degenerate in our discussion, such that the DM dynamics can be captured by an effective single species $A$.  It is also possible to split $\psi_A$ from $\phi_A$ to achieve a more varied phenomenology.  

There is considerable freedom in choosing the masses of $A$ and $B$, though we focus on scales relevant for describing the GC excess:
\be
m_A, m_B \simeq \mathcal{O}(10\text{--}100~\GeV).
\ee
The mass of the $\phi$ is assumed to be small compared to the other scales in the theory, $m_\phi \simeq\mathcal{O}(10~\MeV\text{--}1~\GeV)$, consistent with $\phi$ being a light dark photon that mixes with the SM photon or a light scalar that mixes with the Higgs sector.  We take the couplings of $\phi$ to SM states to be large enough for $\phi$ to stay in thermal equilibrium in the early universe, but small enough to avoid direct detection bounds on $A$ (see \Sec{sec:constraints}).

\begin{figure}[t]
    \centering
    \subfloat[]{{\includegraphics[scale=0.8,trim=-0.2cm 0 0 0]{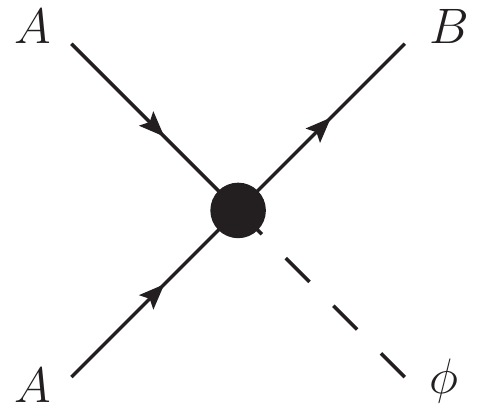}  } \label{fig:interactions:a} }
    \qquad 
    \subfloat[]{{\includegraphics[scale=0.8,trim=-0.3cm 0.1cm 0 0]{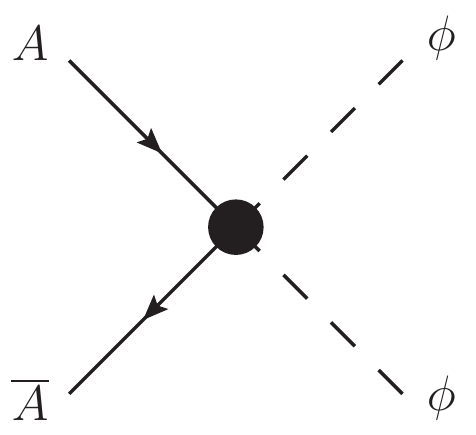} }  \label{fig:interactions:b} }
    \qquad
    \subfloat[]{{\includegraphics[scale=0.8,trim=-0.4cm 0 0 0]{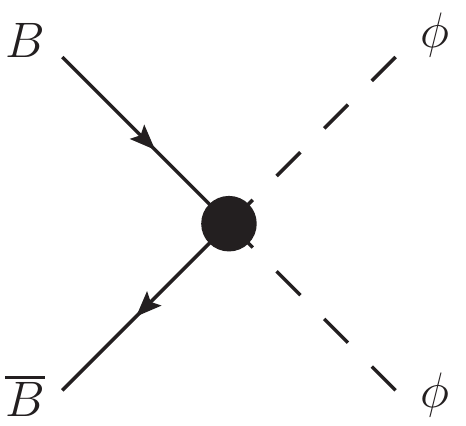}  } \label{fig:interactions:c} }
    \caption{Leading interactions that determine the freeze out of $A$:  (a) semi-annihilation which is also relevant for present-day indirect detection signals, and (b,c) annihilations which ensure depletion of the symmetric components of $A$ and $B$.}
    \label{fig:interactions}
\end{figure}

The key interactions in this scenario are shown in \Figs{fig:interactions}{fig:interactionsBSM}.  The processes in \Fig{fig:interactions} control the freeze-out abundance of $A$, and for simplicity, we assume these processes have velocity-independent cross sections.  The semi-annihilation process in \Fig{fig:interactions:a} has various crossed and conjugate channels of relevance:
\be
AA \to B \phi, \qquad \Abar \Abar \to \Bbar \phi, \qquad A \Bbar \to \Abar \phi, \qquad \Abar B \to A \phi. 
\ee
In order to have an asymmetric DM model with a suppressed symmetric component, we assume that the annihilation processes in \Figs{fig:interactions:b}{fig:interactions:c},
\be
\label{eq:annihilationdiagrams}
A \Abar \rightarrow \phi \phi, \qquad B \Bbar \rightarrow \phi \phi,
\ee
are highly efficient.  When needed for numerical studies, we take the amplitudes $|\mathcal{M}_{A \Abar \ra \phi \phi}| = |\mathcal{M}_{B \Bbar \ra \phi \phi}| = 1$, though smaller couplings lead to similar results.\footnote{\label{footnote:bound}For such large annihilation amplitudes, one might wonder whether $\phi$ exchange could result in additional bound state formation between $A$ and $B$ particles (see, e.g., \Refs{Wise:2014jva,Wise:2014ola,Petraki:2015hla}).  Such bound states could be avoided either by reducing the $\phi$ coupling strength (perhaps resulting in a small residual symmetric DM component) or by ensuring that the would-be Bohr radius of the bound state is larger than the $\phi$ Compton wavelength (see \App{app:model}).  Alternatively, the dominant annihilation process could be unrelated to the $\phi$ particle, as expected if $A$ and $B$ were part of a larger strongly-coupled dark sector with both stable and unstable components.}  In the extreme asymmetric limit with negligible $\Abar$ and $\Bbar$ abundances, the only (thermally-averaged) cross section of relevance is $\langle \sigma_{AA \ra B \phi} v \rangle$.\footnote{When $\phi$ is sufficiently light, the (semi-)annihiliation process can be Sommerfeld enhanced and therefore velocity dependent.  This adds non-trivial temperature dependence to the evolution of the comoving $A$ and $B$ abundances and can change the viable region of parameter space.  We neglect the Sommerfeld effect in this work, though one generically expects it to boost the present day semi-annihiliation rate compared to the rate during DM freezeout.}

\begin{figure}[t]
    \centering
    \subfloat[]{{\includegraphics[scale=0.9,trim=-1.2cm 0 0 0]{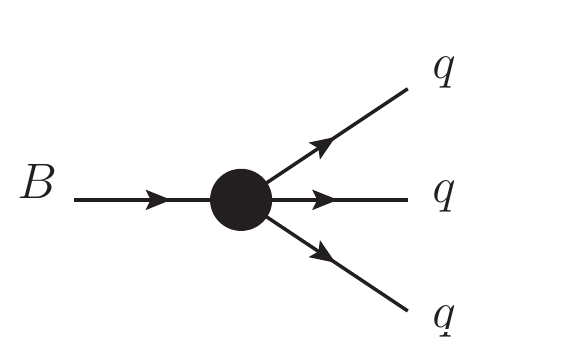} }  \label{fig:interactionsBSM:a} }
    \subfloat[]{{\includegraphics[scale=0.9,trim=-1.2cm 0 0 0]{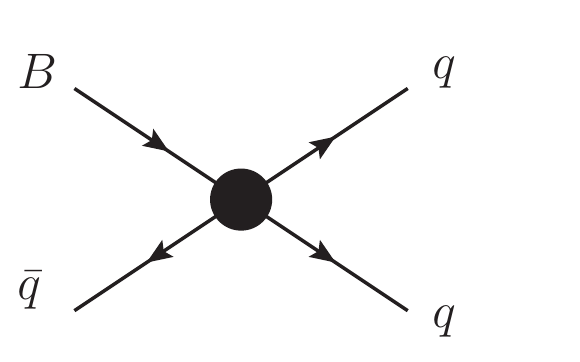}  } \label{fig:interactionsBSM:b} }
    \caption{Leading interactions that determine the asymmetry sharing between DM and the SM: (a) decay of $B$ to SM quarks, and (b) scattering of $B$ with SM quarks.}
    \label{fig:interactionsBSM}
\end{figure}

The processes in \Fig{fig:interactionsBSM} control the sharing of the SM baryon asymmetry with the dark sector.  We denote the total asymmetry by $\eta_{\rm tot}$, which will later be divided into the baryon asymmetry $\eta_b$ and DM asymmetry $\eta_A$.  Assuming $B$ is a fermion with decay modes given by \Eq{eq:decaymodes}, the $B$ decays in \Fig{fig:interactionsBSM:a} are mediated by the operators
\be
\label{eq:decayops}
\frac{1}{\Lambda^2} B u^c d^c d^c, \qquad \frac{1}{\Lambda^2} B c^c b^c b^c,
\ee
where $\Lambda \simeq \mathcal{O}(10\text{--}10^3~\TeV)$ is some new physics scale.\footnote{If $B$ is a scalar, then the decay must include an additional fermion, perhaps a SM lepton or a supersymmetric particle (see, e.g., \Ref{DEramo:2011ec}).  If $B$ is a fermionic bound state of a fermion/boson $\psi_A/\phi_A$ pair, then one should regard \Eq{eq:decayops} as effectively a dimension 7 operator of the form $\psi_A \phi_A u^c d^c d^c$, such that the effective cutoff scale $\Lambda$ would be lower than our benchmark value.}  The $B$ decay width is given parametrically by
\begin{equation}
\label{eq:Blifetime}
\Gamma_B \simeq  \frac{1}{128 (2 \pi)^3}\, m_B \left(\frac{m_B}{\Lambda}\right)^4 \simeq \lp 2.2 \times 10^{-7} \text{ sec}\rp^{-1}  \lp \frac{m_B}{ \text{60 GeV}}\rp^5 \lp \frac{300  \text{ TeV}}{\Lambda}\rp^4.
\end{equation}
This short lifetime ensures that $B$ decays prior to the start of big bang nucleosynthesis (BBN), $T \simeq 10~\MeV$, corresponding to $t \simeq 10^{-2} \, \text{sec}$ \cite{Wagoner:1966pv,Agashe:2014kda}.  The same operators also contribute to the $2 \to 2$ scattering processes involving $B$ shown in \Fig{fig:interactionsBSM:b},  which is the mechanism responsible for the initial sharing of the asymmetry between the SM and the DM in the early universe (see \Fig{fig:Decoupling} below).

As we will see in the next section, there are three temperatures of interest:  the $B$ chemical decoupling temperature $T_D$, the $A/\Abar$ (and $B/\Bbar$) symmetry imbalance temperature $T_I$, and the $A$ freeze-out temperature $T_F$.  While $T_D$ depends sensitively on the scale $\Lambda$, as long as $T_D > \max\{T_I, T_F\}$ then the details of $B$ decoupling are irrelevant for the evolution of the Boltzmann system.  Furthermore, if $T_I \gtrsim T_F$, then the details of the $\Abar$ annihilation are relatively unimportant.  Assuming the hierarchy $T_D > T_I \gtrsim T_F$, the dominant phenomenology of this scenario can be determined by four parameters: 
\be
\{m_A , m_B/m_A, \langle \sigma_{AA \ra B \phi} v \rangle, \eta_{\rm tot} \}.
\ee
Below, we refer to the following benchmark parameters, which yields the desired behavior specified in the introduction with the $B \to cbb$ decay channel:
\be
 \label{eq:benchmark}
m_A = 60 ~\GeV, ~~ m_B/m_A = 1,  ~~ \langle \sigma_{AA \rightarrow B\phi} v \rangle = 3.3 \times 10^{-26} \,  \text{cm}^3/\text{sec}, ~~ \eta_{\rm tot} =  9.4 \times 10^{-11}.
\ee
Note that this benchmark cross section value is comparable to canonical WIMP scenarios.  The proximity of $m_A$ and $m_B$ is because the GC excess fit in \Sec{sec:GC} prefers a higher Lorentz boost factor for $B$.\footnote{In addition,  as shown in \Fig{fig:DMabundance:b}, taking the ratio $m_B/m_A$ to be in the vicinity of 1 gives fine control over the DM abundance for fixed semi-annihilation cross section, making it easier to simultaneously fit the GC excess and achieve the desired DM abundance.}  As shown in \Sec{sec:resultsabundance}, we get reasonable DM phenomenology whenever $m_A \lesssim m_B < 2 m_A$.

One might be surprised that $\eta_{\rm tot}$ in this benchmark is comparable to the present-day baryon asymmetry $\eta_b \simeq 9 \times 10^{-11}$, such that DM only takes around $1/20$ of the asymmetry.  This is unlike usual asymmetric DM scenarios where the total asymmetry $\eta_{\rm tot}$ is roughly twice the baryon asymmetry $\eta_b$ \cite{Petraki:2013wwa,Zurek:2013wia}.  The main reason for the mismatch is that $m_A$ is around ten times heavier than the standard $5$ GeV benchmark for asymmetric DM (i.e.\ the mass that yields the right DM abundance when $\eta_{\rm DM} \simeq \eta_b$).  Also note that the concrete model in \App{app:model} has both a boson and fermion species of $A$, and $A$ has a baryon number $1/2$ (see \Eq{eq:asymmetries} below), further affecting the expected mass relation.

\section{Early Universe Cosmology}
\label{sec:earlyhistory}

\subsection{Initial Conditions}

We start from a primordial asymmetry generated through an unspecified mechanism (see \cite{Petraki:2013wwa,Zurek:2013wia} for a review of options).  The total asymmetry $\eta_{\rm tot}$ is then conserved through the thermal history of the universe by the processes in  \Figs{fig:interactions}{fig:interactionsBSM}.  We use the notation $Y_i = n_i/s$, where $n_i$ is the number density of species $i$ and $s$ is the entropy density.  We can express the total asymmetry as
\be
\label{eq:asymmetries}
\eta_{\rm tot} =   \eta_A + \eta_B + \eta_b, \qquad \eta_A = \frac{1}{2} \, \lp Y_A- Y_{\Abar} \rp, \qquad \eta_B = Y_B- Y_{\Bbar},
\ee
where the factor of $1/2$ accounts for the baryon number of $A$ and $\eta_b$ is the baryon asymmetry (baryons minus anti-baryons).  For ease of discussion, we refer to DM as a single particle $A$.  Because $B$ is a fermion, however, our explicit model in \App{app:model} requires $A$ to be a degenerate fermion-boson system, a fact which is reflected in the equations below.

At  high temperatures, the SM and DM asymmetries are related since the operators in \Eq{eq:decayops} (or their ultraviolet completions) ensure that the two sectors are in chemical equilibrium.    A full analysis of the chemical potentials is presented in \App{app:chemical}.  In equilibrium at a temperature $T$, the asymmetries are:
\begin{eqnarray}
\label{eq:etaA}
\frac{\eta_{A}^\eq}{\eta_{\rm tot}} =\frac{ h_A  \left(  f_{\rm f}(m_A/T)+ f_{\rm b}(m_A/T) \right) }{h_A \left( \, f_{\rm f}(m_A/T)+ f_{\rm b}(m_A/T) \right) + h_B f_{\rm f}(m_B/T) + h_b \,  f_{\rm f}(0)}\,,\\
\label{eq:etaB1}
\frac{\eta_{B}^\eq}{\eta_{\rm tot}} =  \frac{h_B \, f_{\rm f}(m_B/T) }{h_A \left( \, f_{\rm f}(m_A/T)+ f_{\rm b}(m_A/T) \right) + h_B \, f_{\rm f}(m_B/T) + h_b\,  f_{\rm f}(0)}\,, \\
\label{eq:etab}
\frac{\eta_{b}^\eq}{\eta_{\rm tot}} = \frac{h_b \, f_{\rm f}(0)}{h_A \left( \, f_{\rm f}(m_A/T)+ f_{\rm b}(m_A/T) \right) + h_B \, f_{\rm f}(m_B/T) + h_b \, f_{\rm f}(0)}\,,
\end{eqnarray}
where for the explicit model in \App{app:model}, $\{h_A, h_B, h_b \} = \{1/4, \, 1, \, 45/29 \approx 1.6\}$.  (The factor of $45/29$ is familiar from ordinary asymmetric DM scenarios, see e.g.~\cite{Ibe:2011hq}).  The function $f$ is defined in \Eq{eq:fx}; it has the asymptotic behavior $f_{\text{f}}(x) \to 1/6$ for fermions and $f_{\text{b}}(x) \to 1/3$ for bosons as $x \to 0$ (early times) and $f(x) \to 0$ as $x \to \infty$ (late times).

\subsection{Chemical Decoupling of $B$}
\label{sec:chemdecouplingB}

The equilibrium conditions in Eqs.~(\ref{eq:etaA})--(\ref{eq:etab}) hold as long as the interactions between the SM and the $B$ particles in \Fig{fig:interactionsBSM} are active, allowing efficient sharing of the asymmetry.  When both the $B \overline{q} \to qq$ scattering and $B \to qqq$ decay processes go out of equilibrium, then the $B$ particles chemically decouple from the SM.  To estimate the decoupling temperature $T_D$, we compare the rate of $B$ scattering/decay to the Hubble expansion 
\begin{equation}
\label{eq:decoupling}
\max\{n_{q0}^\eq (T_D) \,  \text{exp}(-\mu_q/T_D) \langle \sigma v\rangle_{B \overline{q} \rightarrow qq}, \Gamma_B \} \simeq H(T_D),
\end{equation} 
where the thermally-averaged rate for $Bq \to qq$ scattering is calculated in \App{app:crosssection}, $\langle \sigma v\rangle_{B \overline{q} \rightarrow qq} \sim T^2/\Lambda^4$ in \Eq{eq:sigmaf}, the number density of quarks $n_{q0}^\eq \sim T^3$ is given in \Eq{eq:SMeq},  the chemical potential of quarks $\mu_q$ is defined in \Eq{eq:chempot}, and the $B$ decay width $\Gamma_B$ is given in \Eq{eq:Blifetime}.   The Hubble parameter is
\begin{equation}
 H(T) = 1.66 \sqrt{g_*}\, \frac{T^2}{M_{\rm{Pl}}},
\end{equation}
where $g_*$ is the number of relativistic degrees of freedom at the temperature $T$ \cite{Kolb:1990vq}.  Because $B$ interacts with the SM via the contact operators in \Eq{eq:decayops}, the scattering process are more relevant at early times (high temperatures) while the decay process is more relevant at late times (low temperatures).\footnote{In addition to processes involving on-shell $B$ particles, there are processes mediated by off-shell $B$ particles that can be relevant for chemical decoupling:  $A A \to qqq$ and $A + \overline{q} \to \Abar q q$.  While these are the same order in $\Lambda$ as the processes in \Fig{fig:interactionsBSM}, they have a subdominant effect due to their $2 \to 3$ phase space suppression.}

\begin{figure}[t]
  \centerline{\includegraphics[scale=0.6]{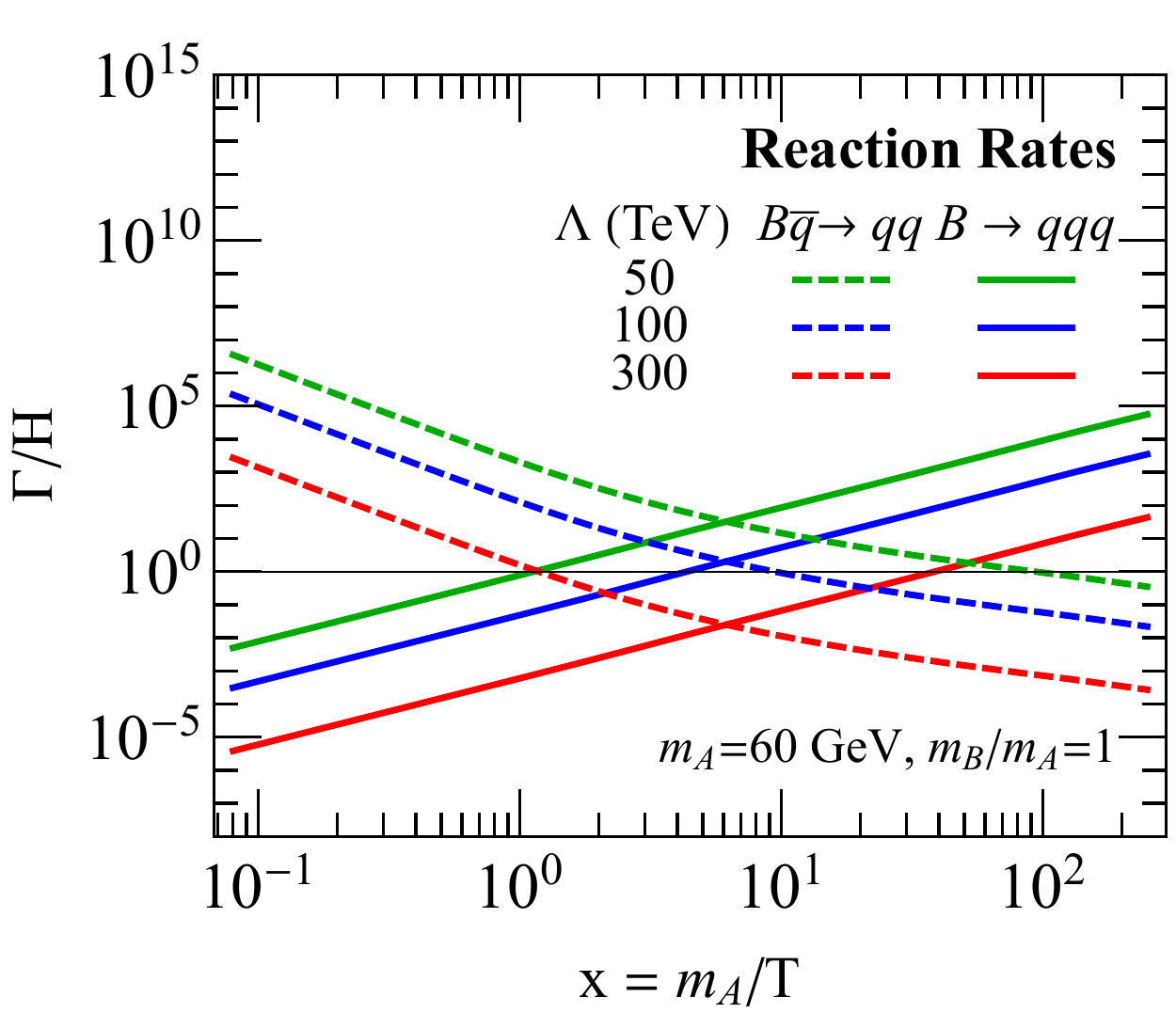}} 
\caption{Comparison of the scattering rate of $B \overline{q} \rightarrow qq$ and the decay rate of $B \rightarrow qqq$ with the Hubble expansion rate. By \Eq{eq:decoupling}, the decoupling temperature $T_D$ occurs when the scattering rate falls below the expansion rate, which in turn sets the initial conditions for the subsequent Boltzmann evolution of $A$ particles.   Chemical equilibrium between $B$ particles and the SM is restored when the $B$ decay rate becomes relevant, though in our benchmark studies, this happens after $A$ particles have already frozen out.} 
 \label{fig:Decoupling}
\end{figure}
 
In \Fig{fig:Decoupling}, we compare the $B$ scattering/decay rates with the Hubble expansion rate for various choices of $\Lambda$, taking the benchmark dark sector parameters in \Eq{eq:benchmark}.  We plot temperatures in terms of $x = m_A/T$.  For $\Lambda \lesssim 50~\TeV$, decoupling never happens since $B$ and the SM remain in chemical equilibrium throughout the early history of the universe.  This tends to erode the DM asymmetry $\eta_A$, since for $T \ll m_A$, $f(m_A/T) \to 0$ in \Eq{eq:etaA}.  Eventually, the semi-annihilation process $AA \rightarrow B \phi$ freezes out, which stops the depletion of $\eta_A$.  While this could lead to a potentially interesting phenomenology, in this paper we focus on larger values of $\Lambda$ where $B$ decouples prior to $A$ freeze out.  Taking the benchmark in \Eq{eq:benchmark} as an example, $\Lambda = 300$ TeV yields $T_D \simeq 55$ GeV, which corresponds to $x \simeq 1.1$ in \Fig{fig:Decoupling}.

When the dark sector decouples from the SM at $T_D$, the asymmetries can be estimated by evaluating Eqs.~(\ref{eq:etaA})--(\ref{eq:etab}) at $T_D$.  If $T_D$ is much higher than $m_{A,B}$, then $f_{\text{f}}(m_{A,B}/T_D) \to 1/6 $ and $f_{\text{b}}(m_{A}/T_D) \to 1/3 $, and the asymmetries at the time of decoupling are:
\be
\frac{\eta_{b,D}}{\eta_{\rm tot}} \to 0.47 , \qquad \frac{\eta_{A,D}}{\eta_{\rm tot}} \to 0.23 , \qquad \frac{\eta_{B,D}}{\eta_{\rm tot}} \to 0.30 .
\ee
For the benchmark in \Eq{eq:benchmark} with $T_D \simeq 55$ GeV, the actual values are
\be
\label{eq:etainitial}
\frac{\eta_{b,D}}{\eta_{\rm tot}} = 0.54, \qquad \frac{\eta_{A,D}}{\eta_{\rm tot}} = 0.17, \qquad \frac{\eta_{B,D}}{\eta_{\rm tot}} = 0.29,
\ee
which is not so different from the $T_D \to \infty$ limit.  

\subsection{Thermal Freeze-out of $A$}
\label{subsec:Afreezeout}

\begin{figure}[t]
    \centering
        \subfloat[]{{\includegraphics[scale=0.55]{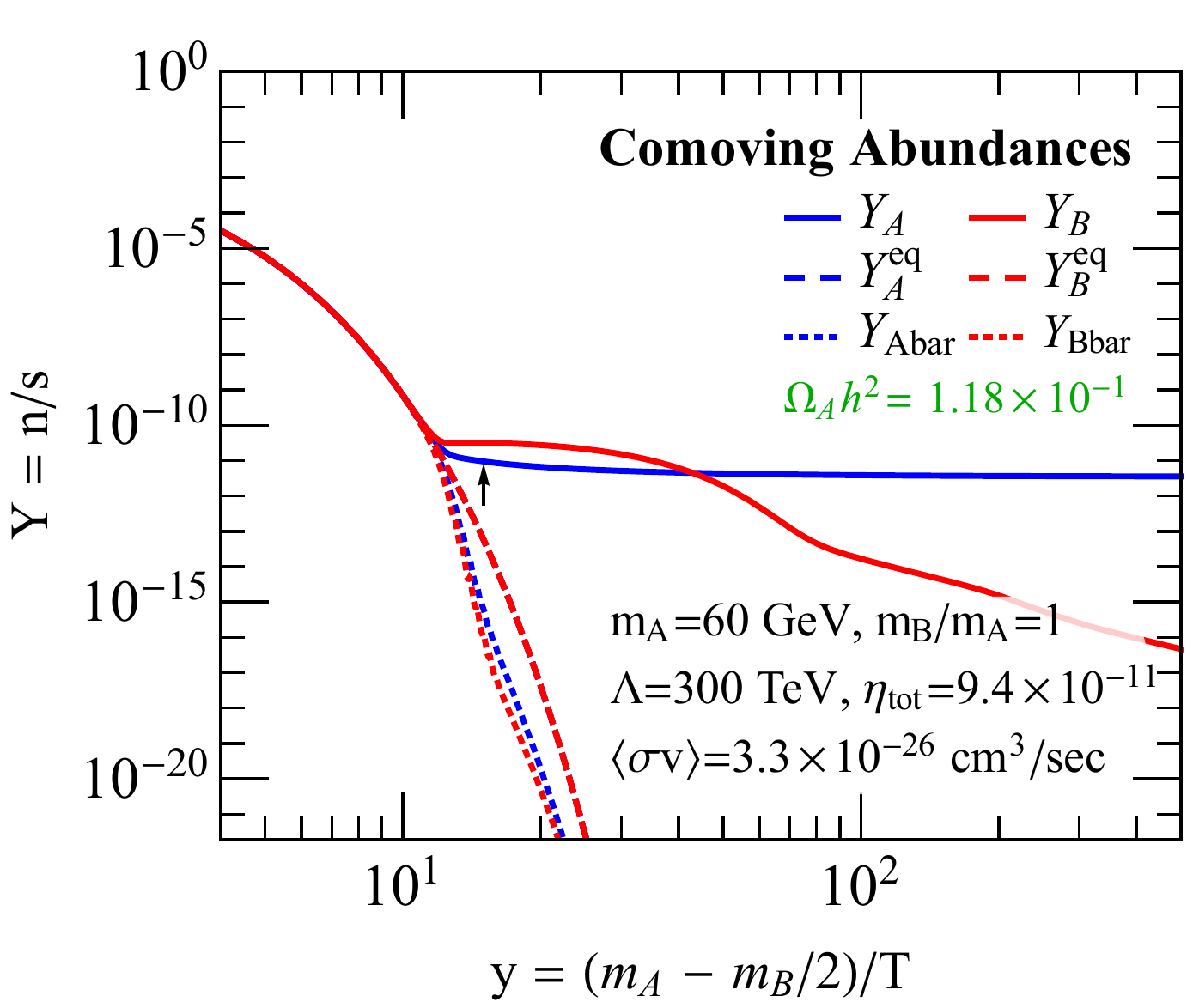}  } \label{fig:DMevolution:a} } 
       \subfloat[]{{\includegraphics[scale=0.56]{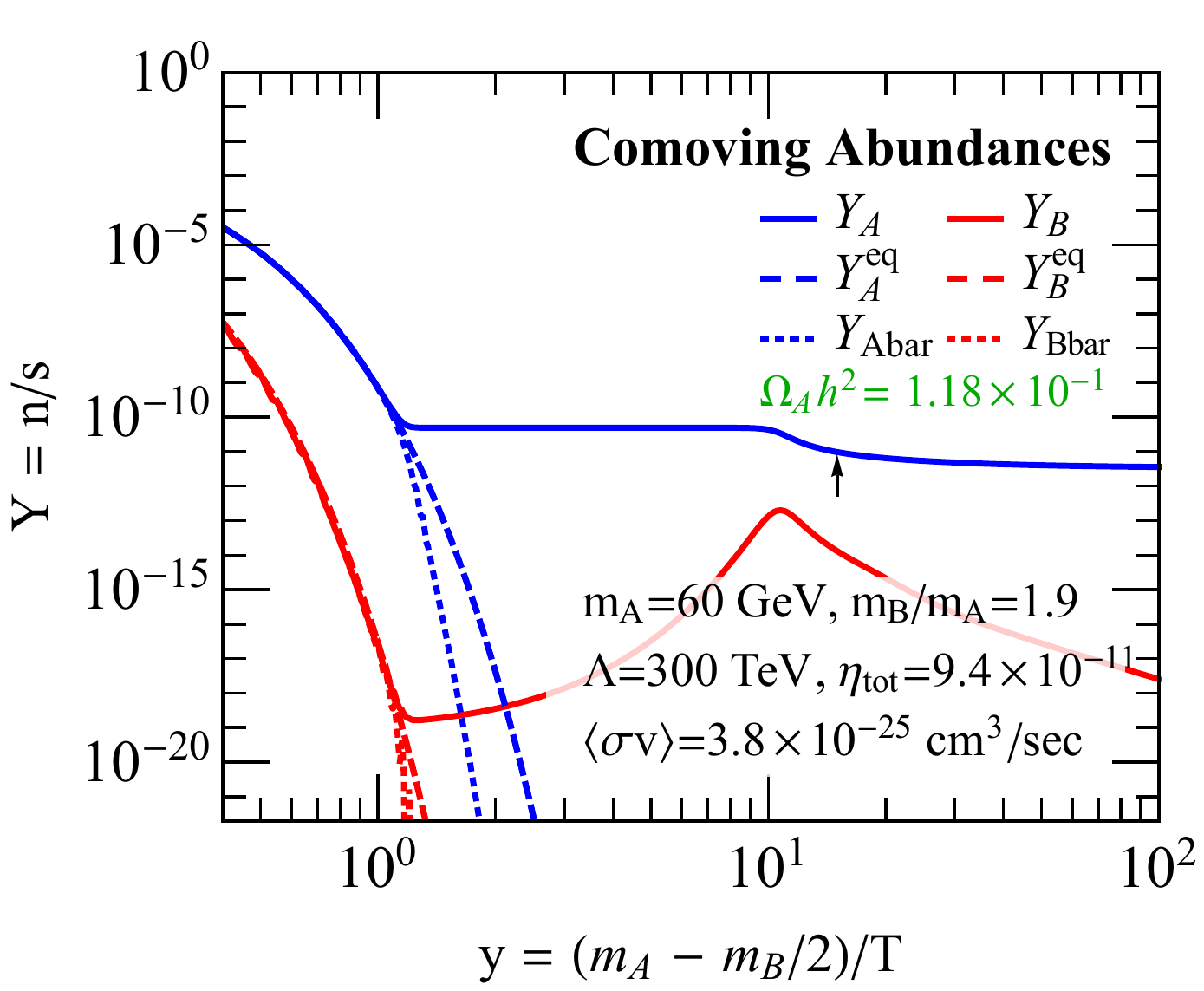}  }\label{fig:DMevolution:b} }
       \caption{Evolution of the comoving abundances of the $\{A,B\}$ system, using (a) the benchmark parameters from \Eq{eq:benchmark} and (b) a more bound-state-like spectrum with $m_B/m_A = 1.9$.  The dashed curves indicate the equilibrium distributions, and the dotted curves indicate the anti-particle distributions. The arrow shows the approximate location of the semi-annihilation freezeout $y_F$. In (b), it is coincidental that the decay of $B$ starts around the same time as $y_F$.}
    \label{fig:DMevolution}
\end{figure}

Once $B$ chemically decouples from the SM, the dark sector follows standard Boltzmann evolution.  The full Boltzmann system is described in \App{app:Boltzmann}, which includes the impact of having initial asymmetries for $A$ and $B$.  There are two relevant dimensionless time variables,
\be
x = \frac{m_A}{T}, \qquad y = \frac{m_A - \frac{1}{2}m_B}{T},
\ee
where the numerator of $y$ was chosen to match the approximate kinetic energy available in the $AA \to B \phi$ process.   Note that $x > y$ whenever $m_B < 2 m_A$.

In \Fig{fig:DMevolution:a}, we show the numerical evolution of the various $Y_X$ for the benchmark scenario in \Eq{eq:benchmark}.  As $x$ increases, the abundances of $A$, $\Abar$, $B$, and $\Bbar$ track the equilibrium distributions until the $A\Abar \to \phi \phi$ and $B\Bbar \to \phi \phi$ annihilation processes freeze out at $x \simeq 20 ~(= x_A)$ and $x \simeq 20 ~m_A/m_B~ (= x_B)$, respectively.\footnote{For a more accurate determination of the freezeout temperature, one should refer to the full asymmetric $x_f$ equation given in \Refs{Iminniyaz:2011yp,Gelmini:2013awa}.  The large annihilation cross sections considered in this paper have an $\mathcal{O}(1)$ impact on $x_A$ and $x_B$.}  At that point, the $\Abar$ and $\Bbar$ abundances rapidly decrease due to the asymmetry.  Assuming $m_B \gtrsim m_A$ as in \Eq{eq:massrelation}, then the effective symmetry imbalance temperature is set by $x_A$, yielding $T_I \simeq m_A / 20$.  Switching to the $y$ variable, the $A A \to B \phi$ semi-annihilation process freezes out around $y_F \simeq 20$, and $T_I \gtrsim T_F$ since $x_A > y_F$.  Eventually, the $B$ particles decay, transferring their asymmetry back to SM baryons.  We are left with a relic DM abundance of $A$ and a baryon asymmetry $\eta_b$, which for this benchmark, match the observed values in our universe.

We now can understand why it is important to take $T_D > T_I$.  If $B$ were in chemical equilibrium with baryons all the way down to $T_I$, then the asymmetry in $A$ would be very suppressed.  Taking \Eq{eq:etaA} with $x_A \simeq 20$, we find $\eta_{A} \simeq 10^{-18}$, which is far too small to obtain a reasonable DM abundance.  For this reason, we need to have $B$ decoupling happen before a large $A/\Abar$ asymmetry develops.  That said, even though $B$ is chemically decoupled from SM baryons at $T_D$, one can still think of the dark sector as being in quasi-equilibrium with baryons out to $T_F$, since any $B$ particles produced will eventually decay to SM baryons. In this way, the asymmetry in the dark sector effectively leaks into the baryon sector via the $AA \rightarrow B \phi$ process, and this leakage stops only when semi-annihilation freezes out.

Perhaps less obvious is why it is important to take $T_I \gtrsim T_F$, which requires $m_B \gtrsim m_A$ since $T_I$ is set by the lower freezeout temperature between $A \Abar \rightarrow \phi \phi$ and $B \Bbar \rightarrow \phi \phi$.  The reason is that if there were a large abundance of $\Bbar$ at later times, then the $A \Bbar \to \Abar \phi$ process would still be active during $A$ freezeout, severely depleting the DM abundance.\footnote{\label{footnote:splitmassBbar}When $A$ is composed of multiple states, for example a fermion $\psi_A$ and a scalar $\phi_A$ as in \App{app:model}, the requirement $m_B \gtrsim m_A$ becomes $m_B \gtrsim \max\{m_{\psi_A},m_{\phi_A}\}$.  This ensures that the processes $\psi_A \Bbar \rightarrow \phi_A^\dagger \phi$ and $\phi_A \Bbar \rightarrow \ov{\psi_A} \phi$ are frozen out earlier than the semi-annihilation process $\psi_A \phi_A \rightarrow B \phi$.}  We will show this effect numerically in \Fig{fig:DMabundance:b} below.

In the extreme asymmetric limit and assuming $T_D \gg T_I \geq T_F$, we can gain an analytic understanding of the DM and baryon abundances.  In this limit, the final $\Abar$ abundance is negligible, so $ \frac{1}{2}\,Y_A(\infty) =  \eta_A(\infty)$ at late times.   After all remaining $B$ particles have decayed, the observed baryon asymmetry today is
\be
\eta_b(\infty) = \eta_{\text{tot}} - \frac{1}{2} Y_A (\infty).  \label{eq:Ybaryon}
\ee
Similarly, the present-day DM relic abundance is
\be
\Omega_{\text{DM}} h^2 = \frac{m_A s_0 h^2}{\rho_C} Y_A (\infty), \label{eq:DMabundance}
\ee
where $s_0 = 2891 ~ \text{cm}^{-3}$ is the entropy density today, and $\rho_C  = 5.15 \times 10^{-6}~\GeV/\text{cm}^3$ is the critical density.  Thus, in this extreme asymmetric limit, finding the baryon asymmetry and the DM abundance reduces to finding $Y_A (\infty)$.

Moreover, in this limit, the value of $Y_A (\infty)$ can be effectively determined by considering just the $AA \to B \phi$ process.  As shown in \App{app:Boltzmann}, the relevant Boltzmann equation is: 
\be 
\frac{\df Y_A}{\df y} = -  \frac{\lambda}{y^2} \langle \sigma_{AA \rightarrow B \phi} v  \rangle \left(Y_A^2 - \lp \eta_{\text{tot}} - \eta_b - \frac{1}{2} Y_A \rp \frac{(Y_A^\text{eq})^2}{Y_B^\text{eq}} \right), \label{eq:BoltYA}
\ee
where $\lambda = s/H$, the Hubble scale $H(T)$ is evaluated at $T = m_A - \frac{1}{2}m_B$, and the equilibrium densities are defined in \Eq{eq:equilibrium}.  We used \Eq{eq:asymmetries} to replace $Y_B$ with $\eta_{\text{tot}} - \eta_b - \frac{1}{2} Y_A$.  Similar to the standard WIMP case  \cite{Kolb:1990vq}, albeit with the replacement $x \to y$, $Y_A$ stops following the equilibrium distribution after freeze-out, and the Boltzmann suppressed term approaches zero.  Using the freeze-out approximation, the solution to \Eq{eq:BoltYA} is
\be
Y_A (\infty) \approx \left( \frac{1}{Y_A (y_F)} + \frac{ \lambda \langle \sigma_{AA \rightarrow B \phi} v \rangle }{y_F } \right)^{-1}. \label{eq:YAinfinity}
\ee
Note that the $(Y_A^\text{eq})^2/Y_B^\text{eq}$ term in \Eq{eq:BoltYA} scales like $e^{-2y}$, compared to the WIMP case with $(Y_A^\text{eq})^2 \sim e^{-2x}$, which explains why freezeout happens at $y_F \simeq 20$ instead of $x_F \simeq 20$.

We can gain further insights into $Y_A (\infty)$ by considering limiting cases.  In the limit of a large $AA \rightarrow B\phi$ cross section, the $1/Y_A (y_F)$ factor drops out, and we recover the familiar WIMP approximation  
\be
 \label{eq:freezeout}
\Omega_{\text{DM}} h^2 = \frac{m_A s_0  h^2}{\rho_C} Y_A (\infty)  \approx  \frac{m_A s_0  h^2}{\rho_C}  \frac{y_F }{ \lambda  \langle \sigma_{AA \rightarrow B \phi} v \rangle   }.
\ee
  Just as for standard WIMPs, $y_F$ has a logarithmic dependence on cross section \cite{Kolb:1990vq}.  In the limit of a small $AA \rightarrow B\phi$ cross section, the $A$ and $B$ particles are effectively decoupled after freeze-out, so the $A$ abundance is set simply by $Y_A (y_F)$, or more accurately, the initial asymmetry at decoupling $ \frac{1}{2} \,Y_A (\infty) \simeq \eta_{A,D}$.  For our benchmark scenario, we end up somewhere in between these two extremes, which is a hybrid of standard WIMP-like behavior (i.e.\ $AA \rightarrow B\phi$ freezeout) and asymmetric DM behavior (i.e.\ initial asymmetries).

\subsection{Results for Dark Matter Abundance}
\label{sec:resultsabundance}

\begin{figure}[t]
    \centering
    \subfloat[]{{\includegraphics[scale=0.6]{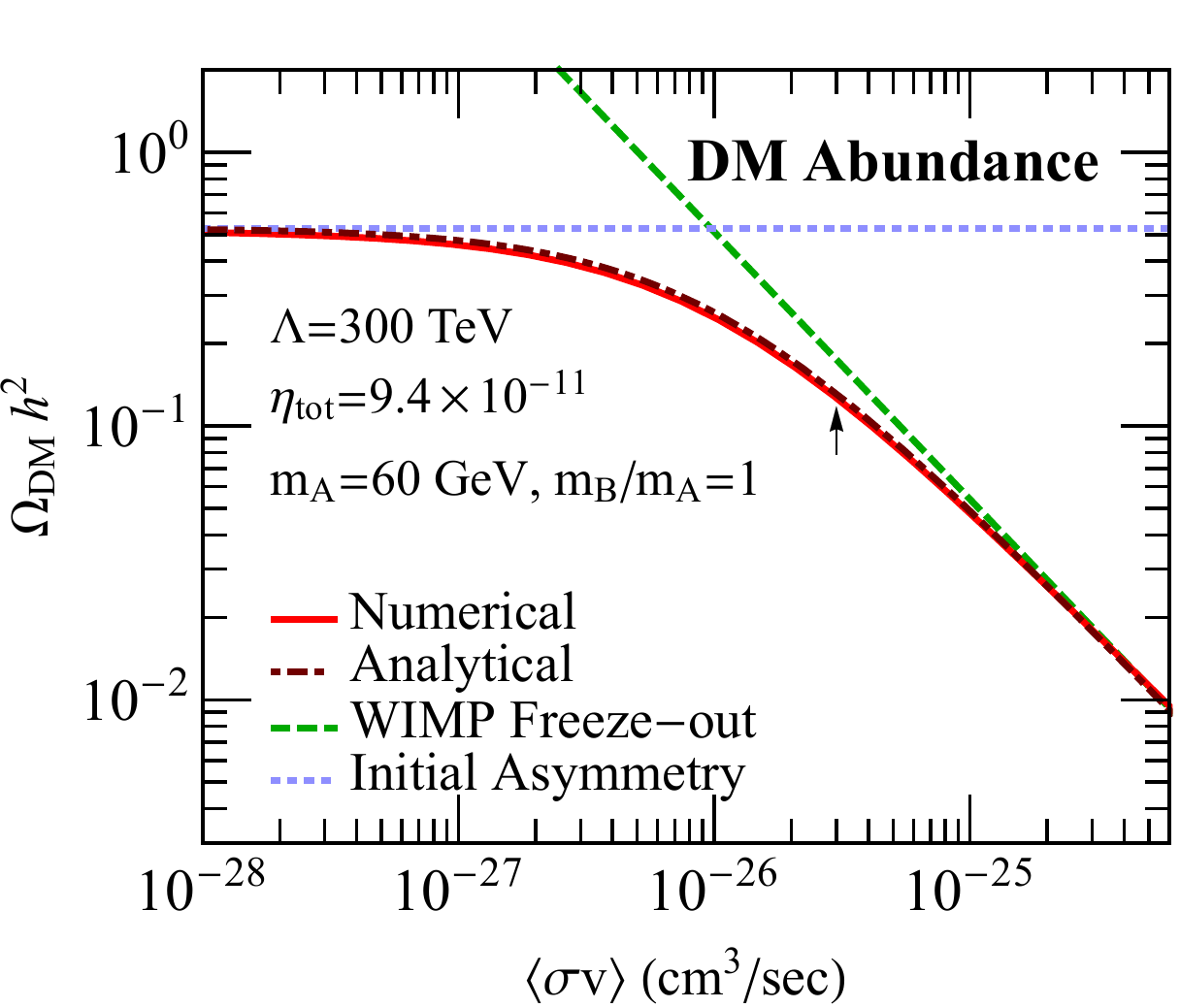}  } \label{fig:DMabundance:a} }
       \subfloat[]{{\includegraphics[scale=0.6]{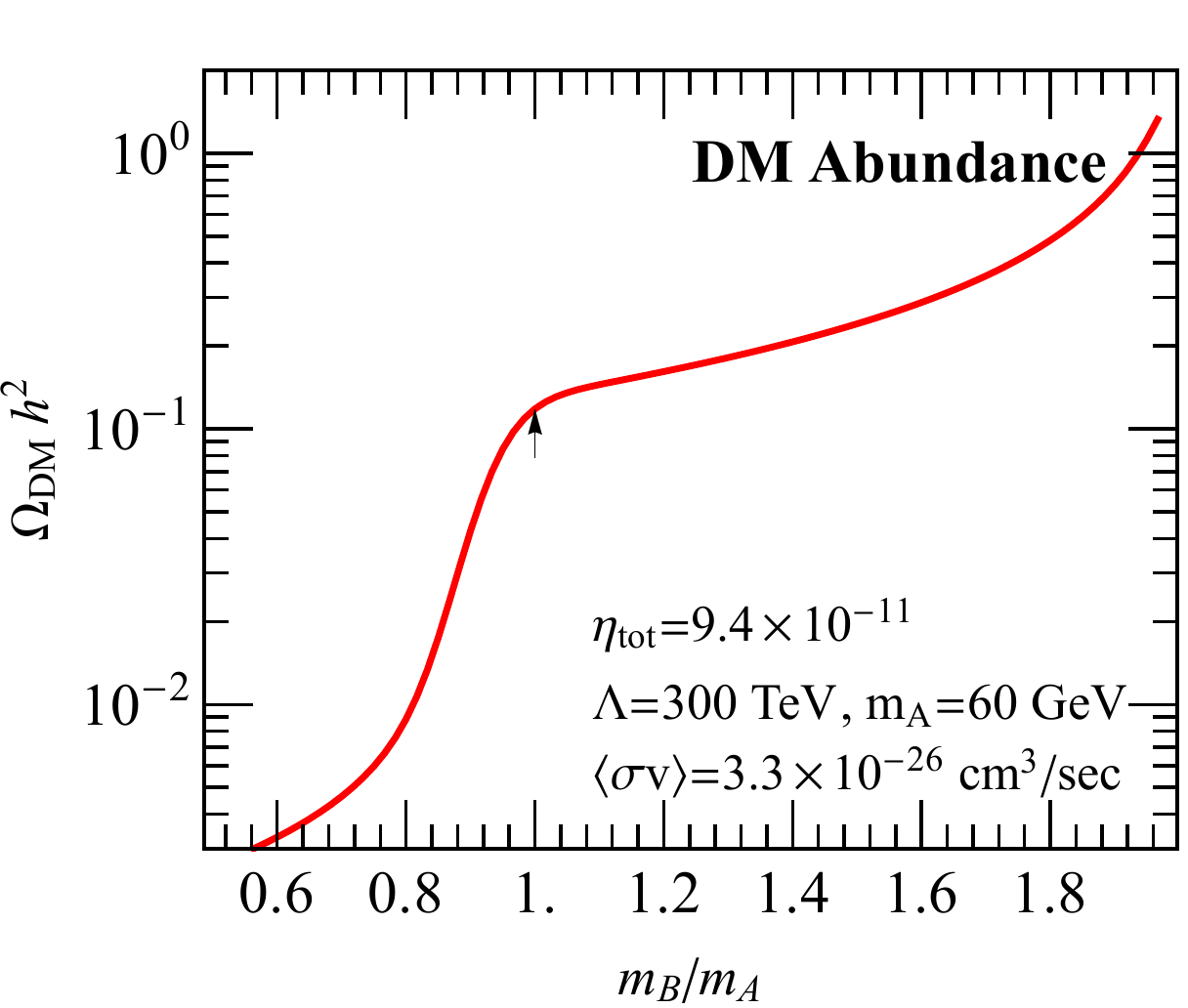}  }\label{fig:DMabundance:b} }
       \caption{(a) DM abundance as a function of the semi-annihilation cross section $\langle \sigma_{AA \ra B \phi} v \rangle$.  The curved dashed line corresponds to the approximation in \Eq{eq:YAinfinity}, while the two straight dashed lines correspond to the extreme freeze-out and asymmetric limits, respectively.   (b)  DM abundance as a function of $m_B/m_A$.  Note the dramatic fall off of $\Omega_{\text{DM}} h^2$ when $m_B < m_A$.   In both plots, unchanged parameters are fixed to the benchmark values from \Eq{eq:benchmark}, and the arrow indicates the benchmark value for the $x$ axis.}
    \label{fig:DMabundance}
\end{figure}

We now present numerical results for the DM abundance and compare to the analytic approximation in \Eq{eq:YAinfinity}.  In \Fig{fig:DMabundance:a}, we show how the DM abundance changes with the $AA \to B \phi$ cross section.  For large values of $\langle \sigma_{AA \rightarrow B\phi} v \rangle$, $\Omega_{\text{DM}}$ asymptotes to the standard thermal freeze-out expectation (albeit with $x \to y$ compared to ordinary WIMPs).  For very small values of $\langle \sigma_{AA \rightarrow B\phi} v \rangle$, the $A$ abundance saturates at the initial asymmetry.

In \Fig{fig:DMabundance:b},   we show how the DM abundance changes as a function of $m_B/m_A$.  For $m_A \lesssim m_B < 2 m_A$, $\Omega_{\text{DM}}$ decreases slowly when decreasing the $m_B/m_A$ ratio.  Once $m_B < m_A$, there is a dramatic drop in $\Omega_{\text{DM}}$, which arises because the $A \Bbar \to \Abar \phi$ process is still active and relevant for determining $Y_A(\infty)$. It is important to note that when $m_B> m_A$, the equilibrium term $(Y_A^\eq)^2/Y_B^\eq $ decreases sharply, faster than the other terms in the Boltzmann system, and therefore setting it to zero after $y_F$ is a valid approximation. This does not hold for $m_B < m_A$, and that is the reason for the difference in behavior between the two regimes.  We are mainly interested in the extreme asymmetric limit where $\Abar$ and $\Bbar$ decouple from the evolution of $Y_A$ (i.e. $T_I \gtrsim T_F$), which is why we focus on the parameter space $m_A \lesssim m_B < 2 m_A$.  Again, for this mass range it is tempting to interpret $B$ as a bound state of two $A$s, though we will not restrict ourselves to that interpretation in this paper.

\begin{figure}[t]
       \centering
       \subfloat[]{{\includegraphics[width = 0.48 \columnwidth]{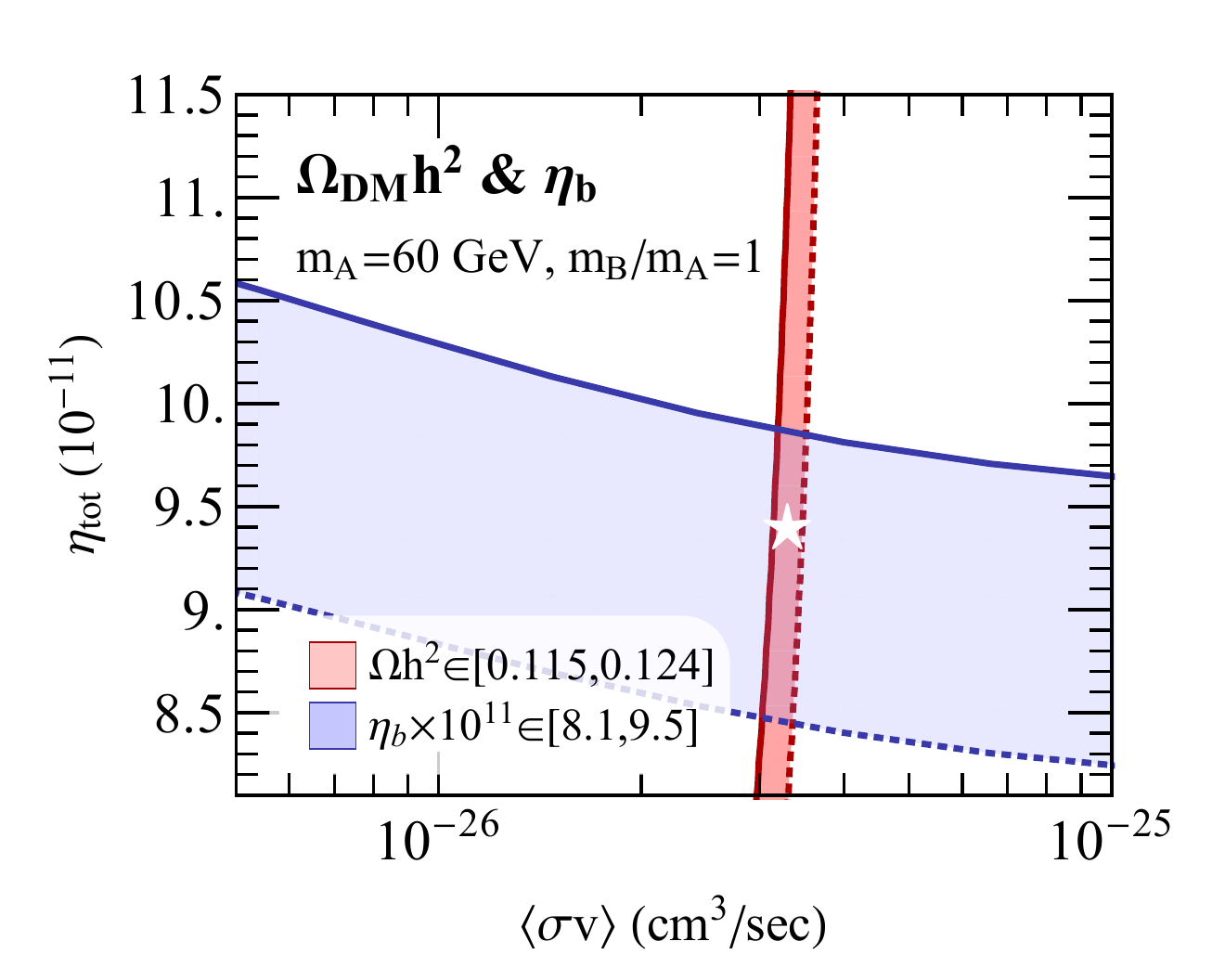}  } \label{fig:allplots:a} }
       \subfloat[]{{\includegraphics[width = 0.48 \columnwidth]{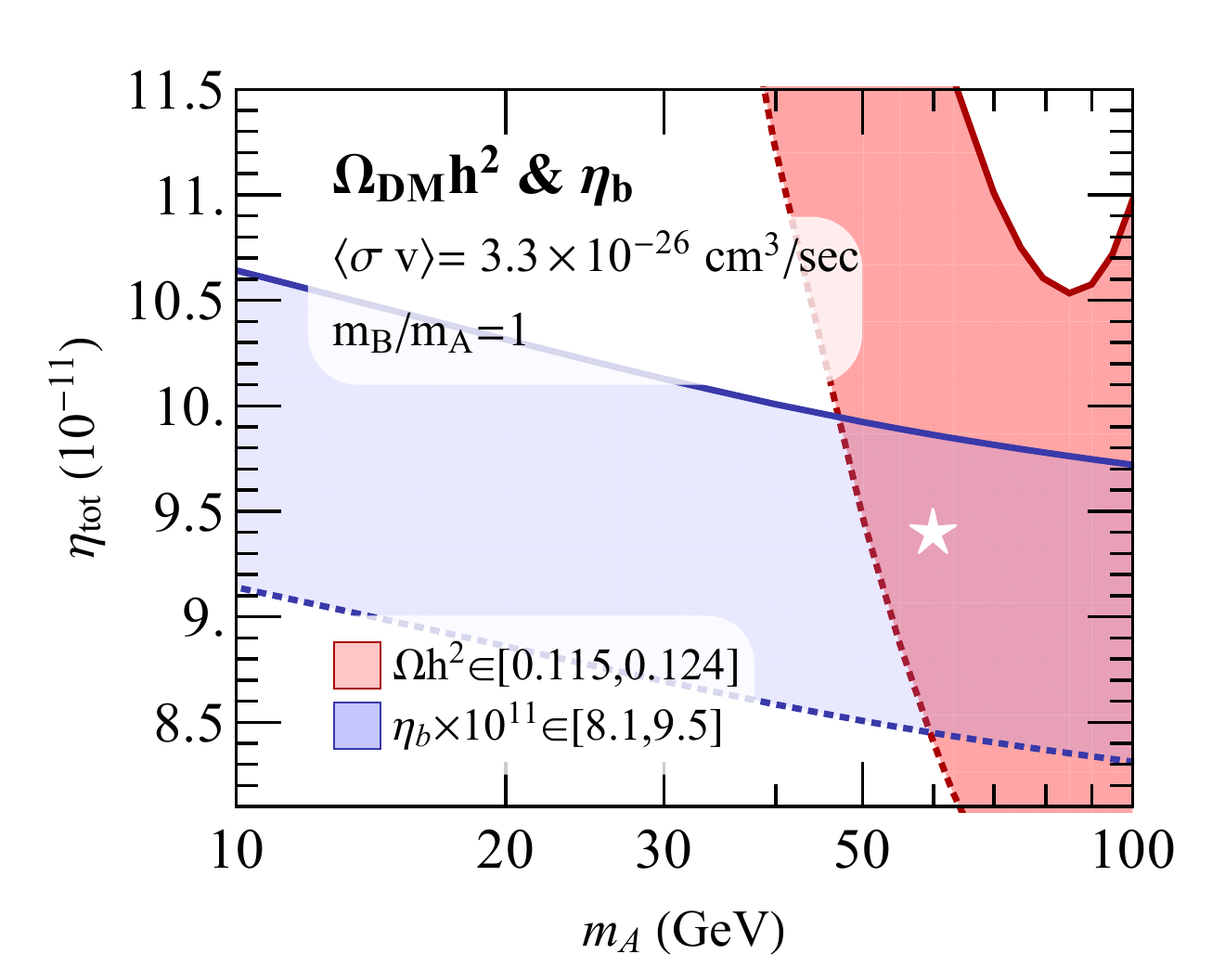} } \label{fig:allplots:b}  }
       
       \subfloat[]{{\includegraphics[width = 0.48 \columnwidth]{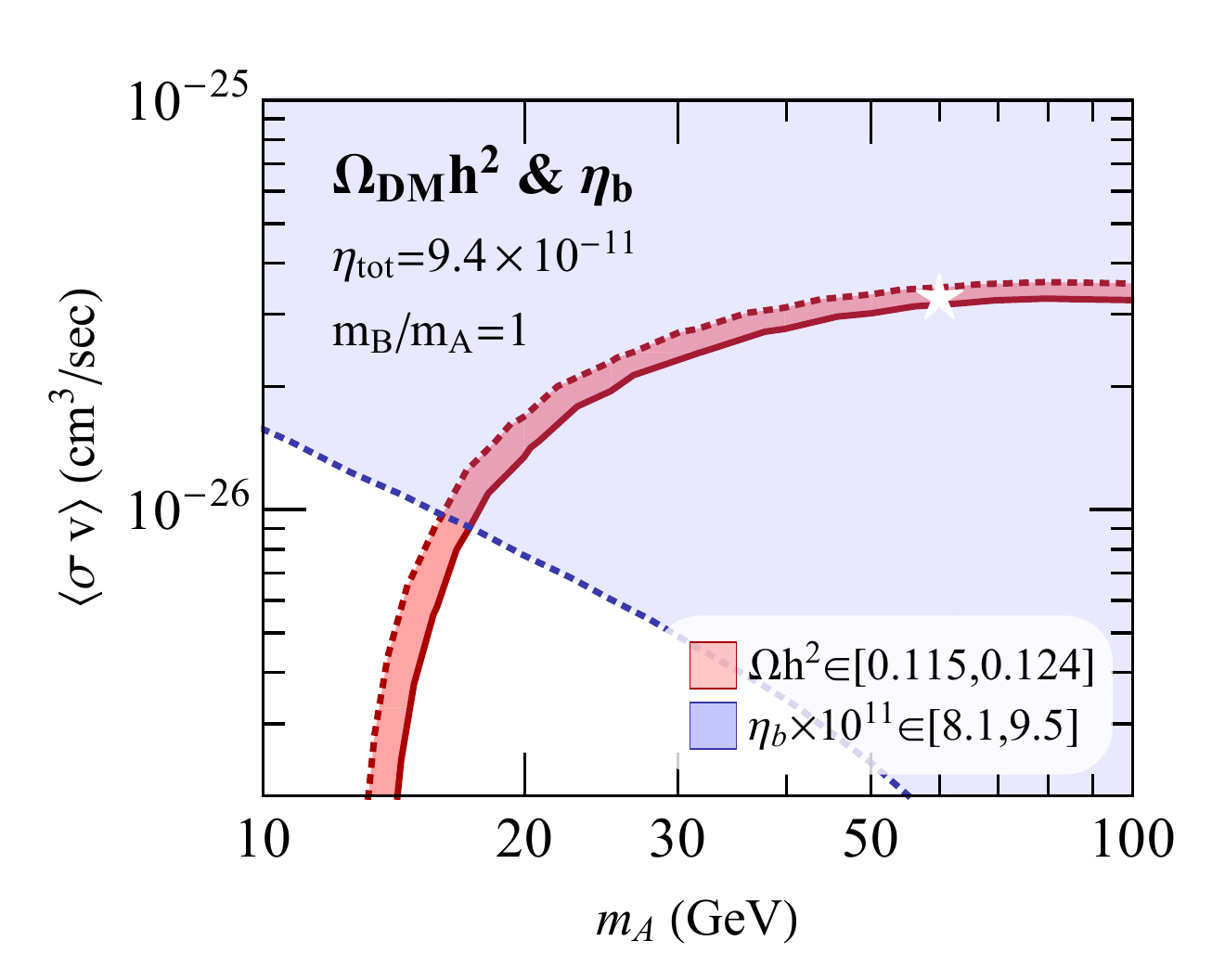} } \label{fig:allplots:c}  }
       \subfloat[]{{\includegraphics[width = 0.48 \columnwidth]{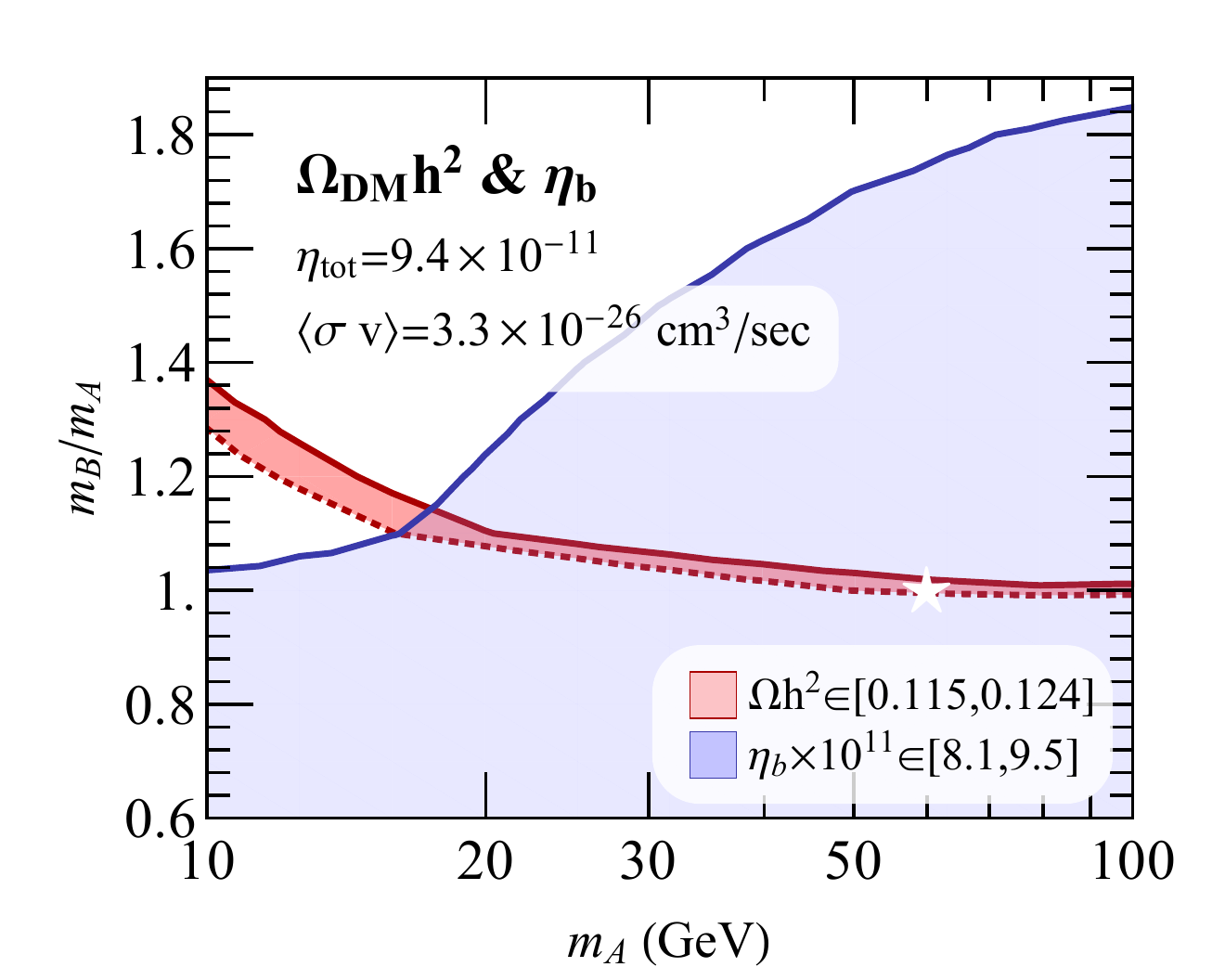} } \label{fig:allplots:d}  }
       \caption{ Regions which satisfy the DM abundance and baryon asymmetry requirements:  $\Omega_{\text{DM}}h^2 \in [0.115, 0.124]$  \cite{Planck:2015xua} (red)  and $\eta_b \in [8.1, 9.5] \times 10^{-11}$ \cite{Agashe:2014kda} (blue).   The star symbol corresponds to the benchmark in \Eq{eq:benchmark}, and from that benchmark value, we sweep (a)   $\langle \sigma_{AA \ra B \phi} v \rangle$ versus $\eta_{\rm tot}$, (b) $m_A$ versus $\eta_{\rm tot}$, (c)  $m_A$ versus $\langle \sigma_{AA \ra B \phi} v \rangle$, and (d) $m_A$ versus $m_B/m_A$.  The dotted (solid) lines indicate the lower (upper) experimental limits.  The sweep in (c) corresponds to the plot shown in the introduction, \Fig{fig:moneyplotintro}.}
\label{fig:allplots}
\end{figure}

In \Fig{fig:allplots}, we show slices through the $\{ m_A, m_B/m_A, \langle \sigma_{AA \ra B \phi} v \rangle, \eta_{\rm tot}\}$ parameter space.  The star marks the benchmark parameters in \Eq{eq:benchmark}.  The red regions correspond to the observed DM abundance ($\Omega_{\text{DM}}h^2 \in [0.115, 0.124]$)  and the blue regions correspond to the observed baryon asymmetry ($\eta_b \in [8.1, 9.5] \times 10^{-11}$).   In \Fig{fig:allplots:a}, the desired DM abundance is achieved with a $AA \ra B \phi$ cross section familiar from the standard WIMP case, with only a small variation with $\eta_{\rm tot}$.  In \Fig{fig:allplots:b}, the required value of $m_A$ also has a weak dependence on $\eta_{\rm tot}$, due to the mild logarithmic dependence of $y_F$ on $m_A$, just as in the WIMP case.  Note that the resulting DM abundance is somewhat correlated with the total asymmetry, since the thermal cross section benchmark is at the transition between the asymmetric regime and the WIMP regime, as shown in \Fig{fig:DMabundance:a}.  

Fixing $\eta_{\rm tot}$, we can highlight the mass and cross section dependence.  In \Fig{fig:allplots:c}, we see the presence of two different regimes in agreement with \Eq{eq:YAinfinity} and \Fig{fig:DMabundance:a}.  In the low cross section limit, the DM abundance is dominated by the initial asymmetry, which is indicated by the vertical behavior at $m_A \simeq 15~\GeV$.  At higher cross sections, $m_A$ has to rise linearly with the cross section in order to obtain the correct abundance using \Eq{eq:freezeout}.  In \Fig{fig:allplots:d}, we also see two regimes which follow from \Fig{fig:DMabundance:b}.  For $m_B/m_A > 1$, the DM abundance band has only a mild dependence on $m_A$.  As $m_B/m_A \to 1$, the number density drops dramatically for fixed cross section, so the value of $m_A$ has to increase to compensate.

\section{Fitting the Galactic Center Excess}
\label{sec:GC}

A number of studies have used the \textit{Fermi} public data to identify a potential excess of gamma rays coming from the GC region \cite{Goodenough:2009gk, Hooper:2010mq, Hooper:2011ti, Abazajian:2012pn, Hooper:2013rwa, Gordon:2013vta, Huang:2013pda, Macias:2013vya, Abazajian:2014fta,  Daylan:2014rsa, Zhou:2014lva, Calore:2014xka}.  The origin of this excess is as-of-yet unknown, but a tantalizing possibility is that this a signature of DM annihilation, though astrophysical explanations are also plausible \cite{2011JCAP...03..010A,Gordon:2013vta,Macias:2013vya,Abazajian:2014fta,Yuan:2014rca,Carlson:2014cwa,Petrovic:2014uda}.   The DM interpretation is bolstered by the fact that the needed annihilation rate is consistent with that of a thermal WIMP, though there is recent evidence that the GC excess is better fit by a population of unresolved point sources \cite{Lee:2015fea,Bartels:2015aea}.  In any case, typical asymmetric DM models do not predict this kind of indirect detection signal, though, unless there is a residual symmetric component (see e.g.~\cite{Hardy:2014dea, Bell:2014xta}).  Here, however, the semi-annihilation process $AA \to B \phi$ followed by $B$ decaying to hadrons can give rise to an interesting gamma ray signal.  In this way, the GC excess could be connected to the process of asymmetry sharing in the early universe.

The decays of $B$ are prompt on cosmological time scales, so one can think of the $AA \to B \phi$ process as being a one-step cascade decay \cite{Pospelov:2007mp,ArkaniHamed:2008qn,Nomura:2008ru,Mardon:2009rc} (see also \cite{Martin:2014sxa, Abdullah:2014lla,Ko:2014gha,Elor:2015tva}) where the $B$ subsequently decays to three quarks.  Recall from \Eq{eq:decaymodes} that our benchmark decay modes are $B \to udd$ and $B \to c b b$, though other flavor combinations are equally plausible.  These quarks hadronize, and the resulting gamma ray spectrum comes primarily from neutral pions which decay via $\pi^0 \ra \gamma \gamma$.  The decays of $\phi$ are model dependent and may contribute to the gamma ray signal as well.  For concreteness, we assume that $\phi$ dominantly decays to $\mu^+ \mu^-$,\footnote{\label{footnote:phimass}This is consistent with $\phi$ being a Higgs portal scalar with $m_\phi \simeq 250~\MeV$.  Of course, for larger $\phi$ masses, the $\phi \to \pi^0 \pi^0$ channel opens up, which yields an additional source of prompt gamma rays.  We take the Higgs portal with $m_\phi \simeq 250~\MeV$ as a benchmark when discussing CMB bounds in \Eq{eq:feffcalc} and direct detection bounds in \Eq{eq:directdetectionbound}, though strictly speaking, for such low $\phi$ masses, one should also account for additional $\phi$-mediated bound states of $A$ and $B$.  In the extended model of \App{app:model}, we use axion-portal-like couplings, where $\phi$ is replaced by a scalar/pseudoscalar pair, in which case there is more flexibility to raise the $\phi$ mass.} which only results in a small contribution to the gamma ray spectrum from final state radiation (FSR).  For simplicity, our study ignores these possible FSR photons as well as photons from inverse Compton scattering or bremsstrahlung.

To determine the gamma ray spectrum in the $B$ rest frame, we use \pythia 8.185 \cite{Sjostrand:2007gs} to construct a color singlet three quark final state, uniformly filling out the allowed three-body phase space.\footnote{A more accurate analysis would take into account the matrix element of the $B \to qqq$ decays, but we use flat three-body phase space to remain agnostic about the Lorentz structure of the decay operator.}  We then use the default hadronization and decay model in \pythia\ to obtain the gamma ray spectrum $\df N / \df E_{\rm rest}$ after letting all unstable hadrons decay.  To go to the $AA \to B \phi$ rest frame, we boost the $B$ particle by the gamma factor
\begin{eqnarray}
\gamma = \frac{ (2 m_A )^2 - m_\phi^2 + m_B^2}{4 m_A  m_B}, \qquad \beta = \sqrt{1 - \frac{1}{\gamma^2}},
\end{eqnarray}
taking $m_\phi = 0$ for simplicity.  The resulting gamma ray spectrum is given by (see, e.g., \cite{Detmold:2014qqa})
\be
\label{eq:boost}
\frac{\df N}{\df E} = \frac{1}{2 \beta \gamma} \int_{E/(\gamma (1+\beta))}^{E/(\gamma (1-\beta))}  \frac{\df E_{\rm rest}}{E_{\rm rest}} \frac{\df N}{\df  E_{\rm rest}}.
\ee

The produced flux of gamma rays as seen on earth is
\be
\frac{\df^2 \Phi_\gamma}{\df \Omega \, \df  E} = \frac{r_\text{sun}}{8 \pi m_A^2 } \, \frac{\df N}{\df E} \, J_\text{norm} \, \langle \sigma_{AA \ra B \phi} v\rangle_\text{eff}.
\ee
The $J$ factor is the integral along the line of sight of the DM density.  We adopt the normalization of \Ref{Calore:2014xka} where the region of interest (ROI) is $| l | \leq 20^\circ$ and $2^\circ \leq |b| \leq 20^\circ$:  
\be
J_{\text{norm}}= \frac{\int_\text{ROI} \df \Omega \, J(l, b)}{\int_\text{ROI} \df \Omega} = 2.06 \times 10^{23}~\GeV^2/\text{cm}^5.
\ee
The effective cross section accounts for the possibility that $A$ comprises only a fraction of the total DM density:
\be
 \label{eq:sigmaeff}
\langle \sigma_{AA \ra B \phi} v\rangle_\text{eff} = \langle \sigma_{AA \ra B \phi} v\rangle \left( \frac{\Omega_A}{\Omega_{\text{DM}}} \right)^2.
\ee
Written this way, the resulting gamma ray spectrum depends on three parameters
\be
\label{eq:threepara}
\{ m_A, m_B/m_A,  \langle \sigma_{AA \ra B \phi} v\rangle_\text{eff} \},
\ee
and the choice of $B$ decay channel.

\begin{figure}[t]
    \centering
    \subfloat[]{{\includegraphics[width = 0.48 \columnwidth]{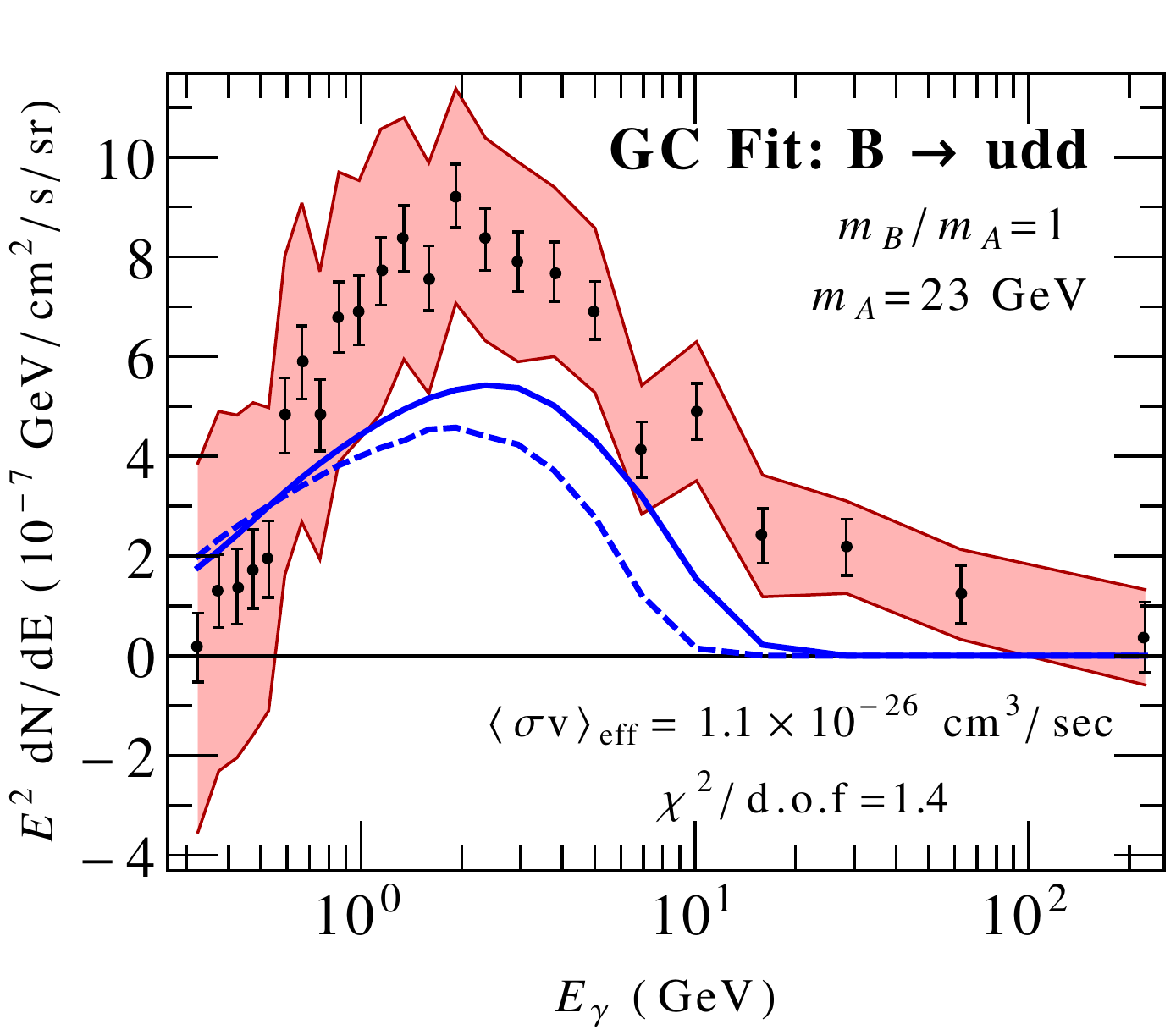}  } \label{fig:annihilationfit:a} }
       \subfloat[]{{\includegraphics[width = 0.48 \columnwidth]{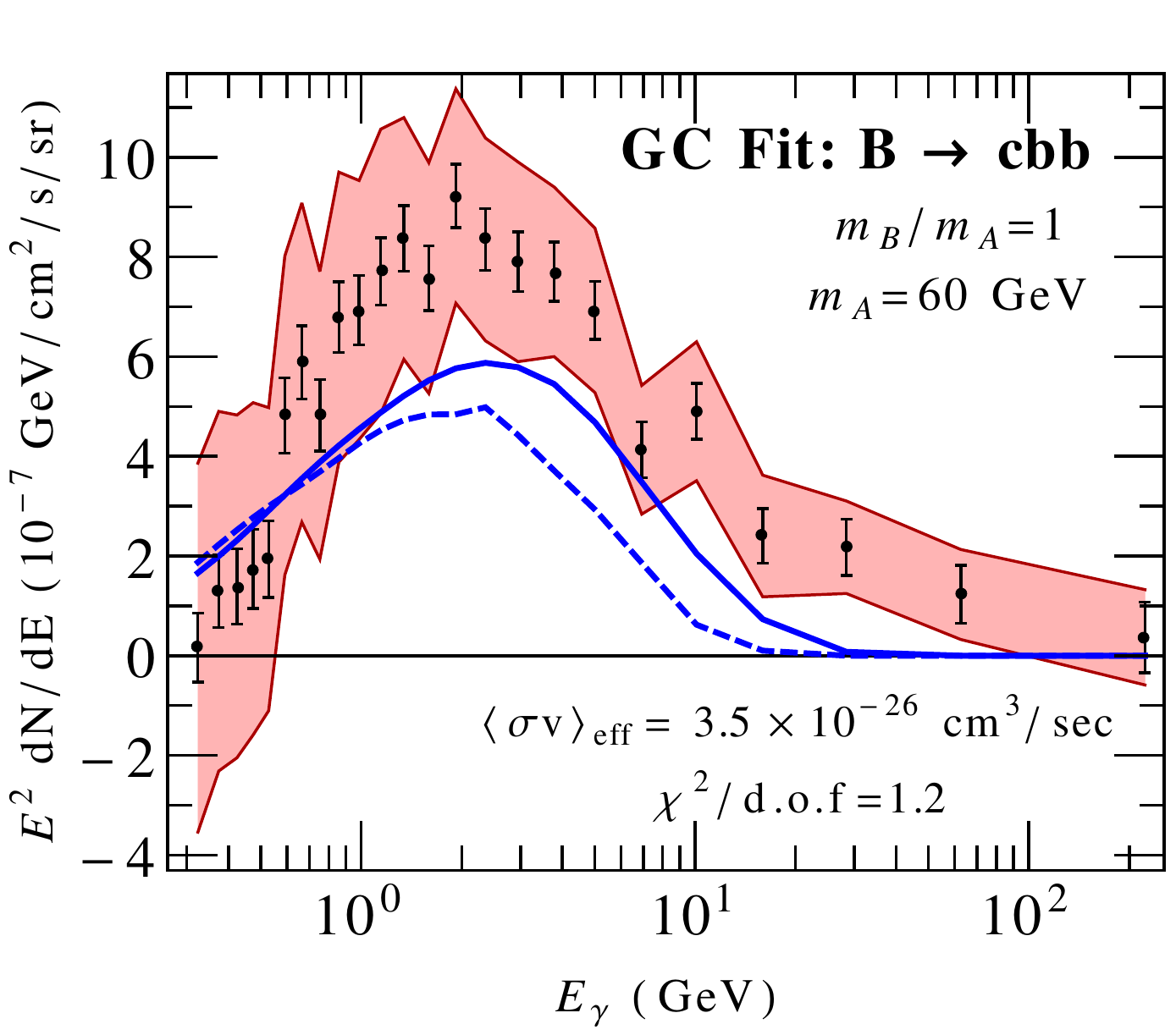} }  \label{fig:annihilationfit:b} }
       \caption{Example fits to the GC excess from $AA \to B \phi$ semi-annihilation with (a) $B \rightarrow udd$ decay and (b) $B \rightarrow cbb$ decay, following the analysis of \Ref{Calore:2014xka}.  The black error bars are statistical uncertainties while the red band represents correlated systematic uncertainties.  The parameters of the model are $m_A$, $m_B/m_A$, and $\langle \sigma_{AA \rightarrow B \phi} v\rangle_\text{eff}$, which makes the number of degrees of freedom $24 - 3 = 21$. The effective cross section $\langle \sigma v\rangle_\text{eff} $ is defined by \Eq{eq:sigmaeff}. The solid (dashed) lines are gamma ray spectra after (before) boosting from the $B$ rest frame to the $AA$ rest frame.}
    \label{fig:annihilationfit}
\end{figure}

To fit the GC excess, we use the procedure outlined in \Ref{Calore:2014xka}.  Adopting the same notation as \Ref{Elor:2015tva}, the chi-squared for a given parameter point is
\be
\chi^2 = \sum_{ij}  \lp E^2 \frac{\df N}{\df E} _\text{i, model} - E^2 \frac{\df N}{\df E}_\text{i, data}  \rp C_{ij}^{-1} \lp E^2 \frac{\df N}{\df E}_\text{j, model} - E^2 \frac{\df N}{\df E}_\text{j, data}   \rp\,,
\ee
where $C_{ij}^{-1}$ is the inverse covariance matrix, obtained from \Ref{Calore:2014xka}.  We show example fits (reasonably close to the best ones) of the photon spectrum in \Fig{fig:annihilationfit}, for both the $B \to udd$ and $B \to c b b$ channels.   Because the photons from heavy quark decays are softer than for light quarks, obtaining a good fit for the $B \to c b b$ channel requires a higher mass than for the $B \to udd$ channel.  This is consistent with the observation for standard WIMP scenarios that $b$ quark final states require higher DM masses than $\tau$ lepton final states \cite{Daylan:2014rsa}.   Because we are considering parameter points for which $m_A \simeq m_B$, the effect of the boost is mild, pushing the peak of the (energy-squared-normalized) photon spectrum to slightly higher values.  Consistently with previous work, the best fitting (effective) cross section is close to the expected WIMP thermal cross section.

\begin{figure} [t]     
       \subfloat[]{{\includegraphics[height= 7cm, trim={0 0cm 0 0},clip]{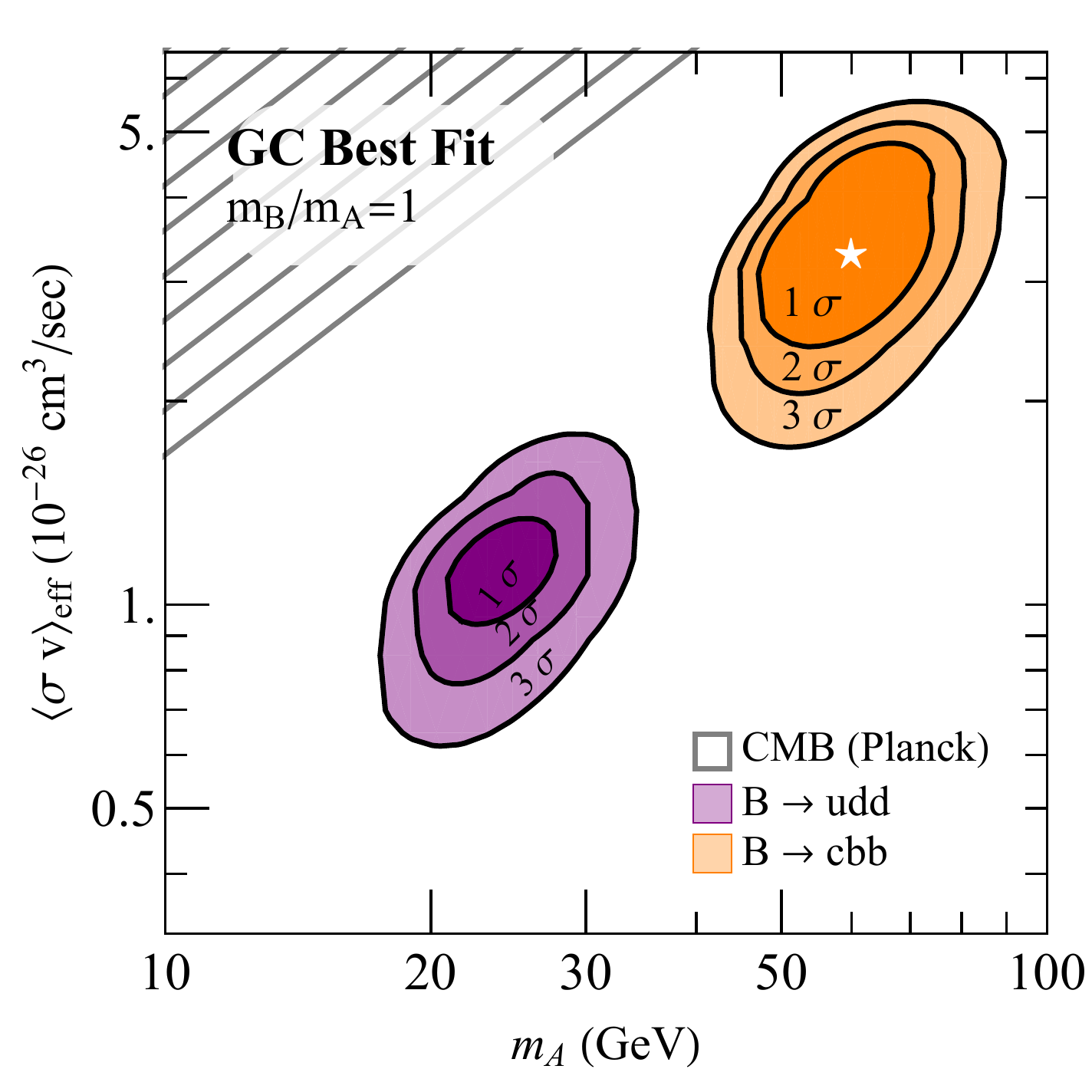} } \label{fig:allplots2:a}   }
       \subfloat[]{{\includegraphics[height= 7cm]{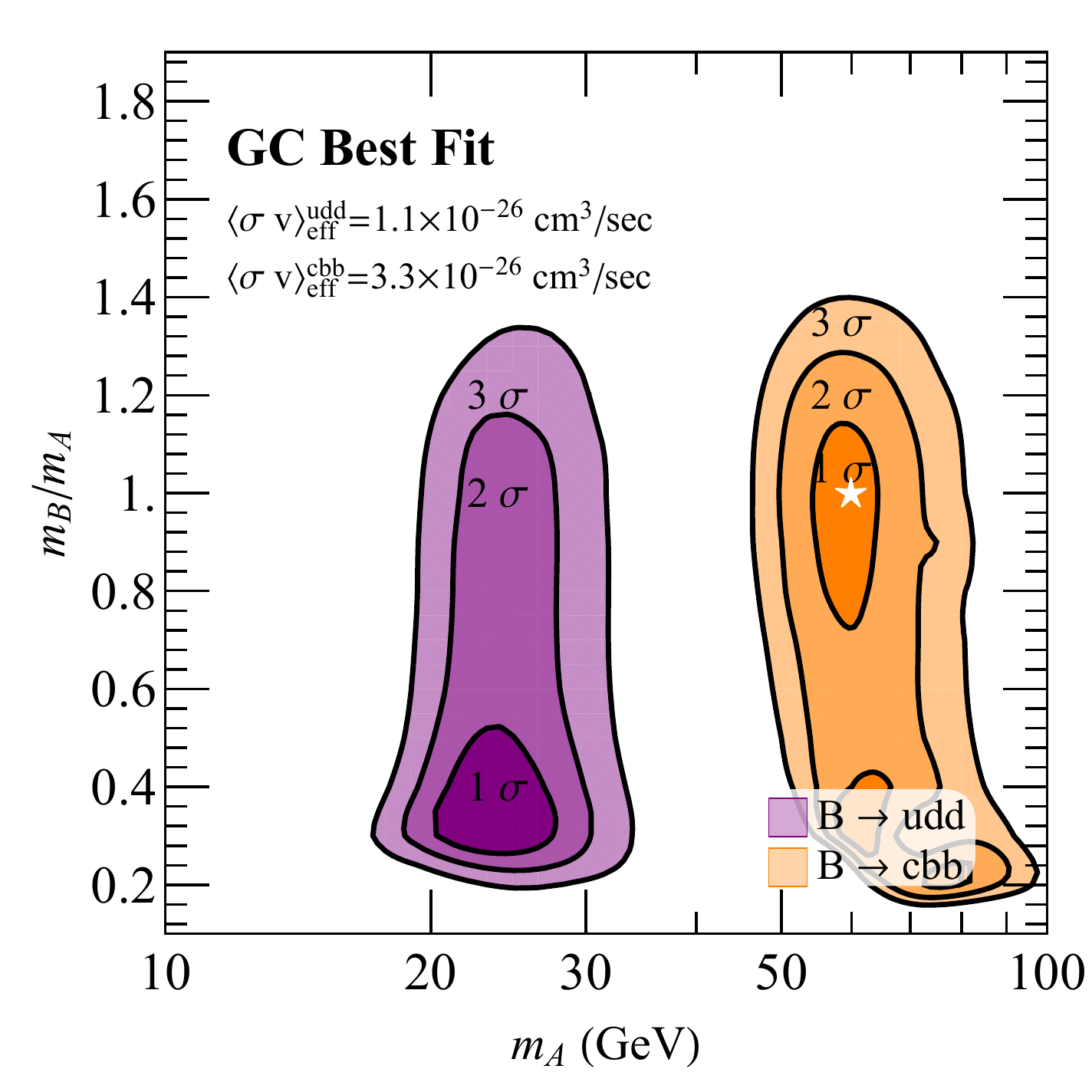} } \label{fig:allplots2:b}  }
       \caption{
      Fits of the GC excess for the $B \to udd$ (purple) and $B \rightarrow cbb$ (orange) channels. The best fit regions correspond to regions of 1, 2, and 3 sigma off the best fit $\chi^2$.  The star symbol corresponds to the benchmark \Eq{eq:benchmark}. In (a), we fix the mass ratio $m_B/m_A$ and then vary the DM mass $m_A$ and effective cross section $\langle \sigma_{AA \ra B \phi} v\rangle_\text{eff}$, while in (b) we fix the effective cross section (to different values in the two channels) and vary the DM mass and mass ratio.  We do not show the CMB limits in (b) as those only constrain lower masses.  Note that the GC best fit point does face potential AMS-02 antiproton bounds (see \Eq{eq:antiprotonbounds}).  The sweep in (a) corresponds to \Fig{fig:moneyplotintro}.}
    \label{fig:allplots2}
\end{figure}

In \Fig{fig:allplots2}, we show the parameter regions that give the best fit to the GC excess in the  $B \to udd$ (purple) and  $B \rightarrow cbb$ (orange) channels.  In each plane, we first find the best fit $\chi^2$, and then show $1$, $2$, and $3$ standard deviation contours for $\Delta \chi^2$.  In \Fig{fig:allplots2:a}, we show the $m_A$ versus $ \langle \sigma_{AA \ra B \phi} v \rangle_\text{eff}$ plane, leaving $m_B/m_A$ fixed, showing that the best fit regions tend to have WIMP-like cross sections.   In \Fig{fig:allplots2:b}, we show the $m_A$ versus $m_B/m_A$ plane, leaving $\langle\sigma_{AA \ra B \phi} v\rangle_\text{eff}$ fixed.  The value of $m_B/m_A$ determines the boost of $B$, and can be used to fine tune the gamma ray spectrum.  Larger boosts (smaller mass ratios) are somewhat preferred by the fit.  

In \Fig{fig:allplots2:a}, we have superimposed the CMB limits discussed below in \Eq{eq:CMB}.  These limits are largely independent of $m_B$ and do not constrain the parameters in \Fig{fig:allplots2:b}.  As noted in \Refs{Planck:2015xua,Kong:2014haa,Cline:2015qha}, models that tend to fit the GC excess are in some tension with the CMB limits, though there is still viable parameter space.  The CMB limits are proportional to a parameter $f_\text{eff}$ that quantifies the fraction of energy that goes into electrons and photons.  We compute this parameter below in \Eq{eq:feffcalc} and find $f^B_\text{eff} = 0.17$ if we only account for the $B$ decay products, though we use the more conservative value $f_\text{eff} = 0.24$ which assumes $\phi \to \mu^+ \mu^-$.

\begin{figure}[t]
   \centering
       \subfloat[]{{\includegraphics[height= 7cm, trim={0 0 0 0},clip]{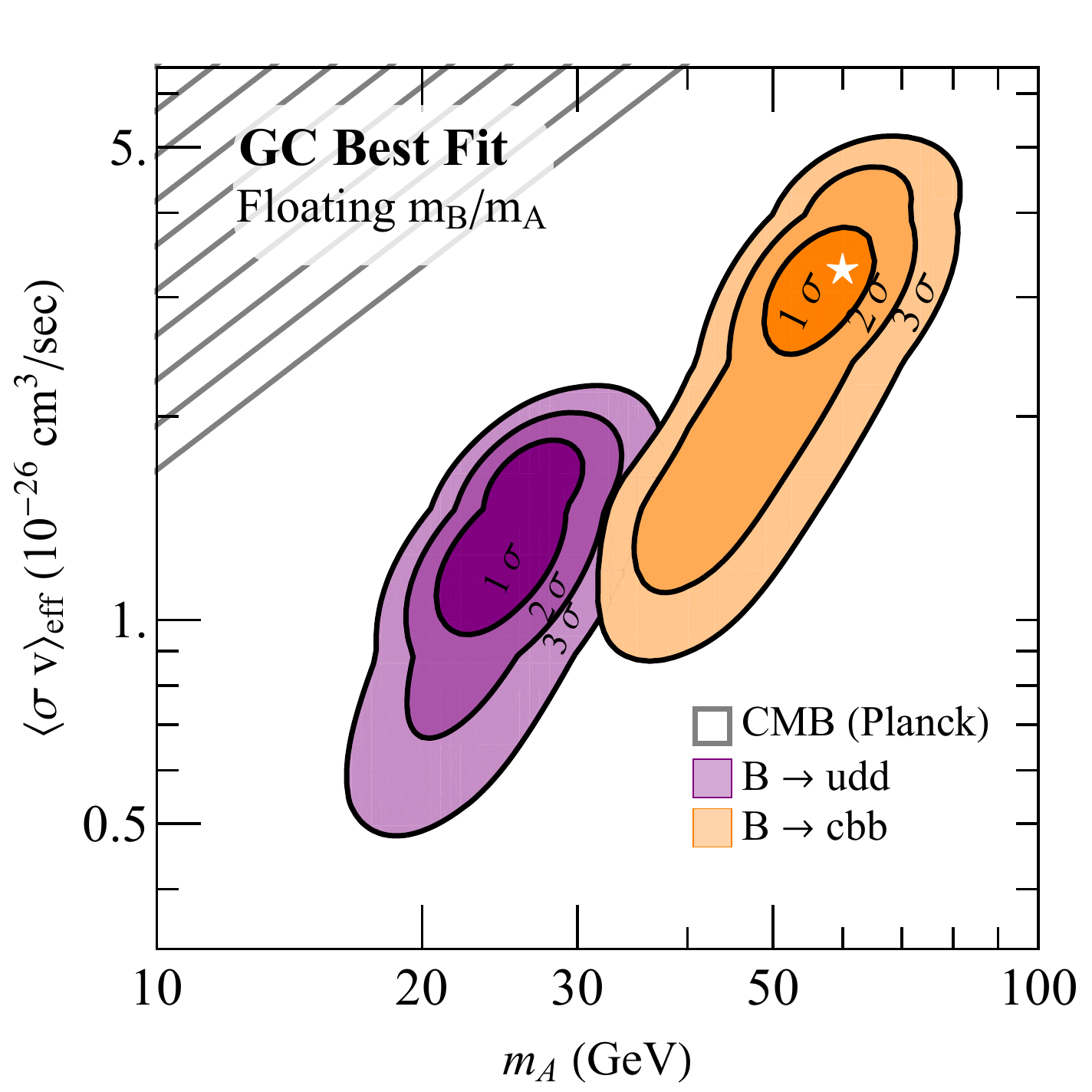} } \label{fig:floatingfits:a}   }
       \subfloat[]{{\includegraphics[height= 7cm]{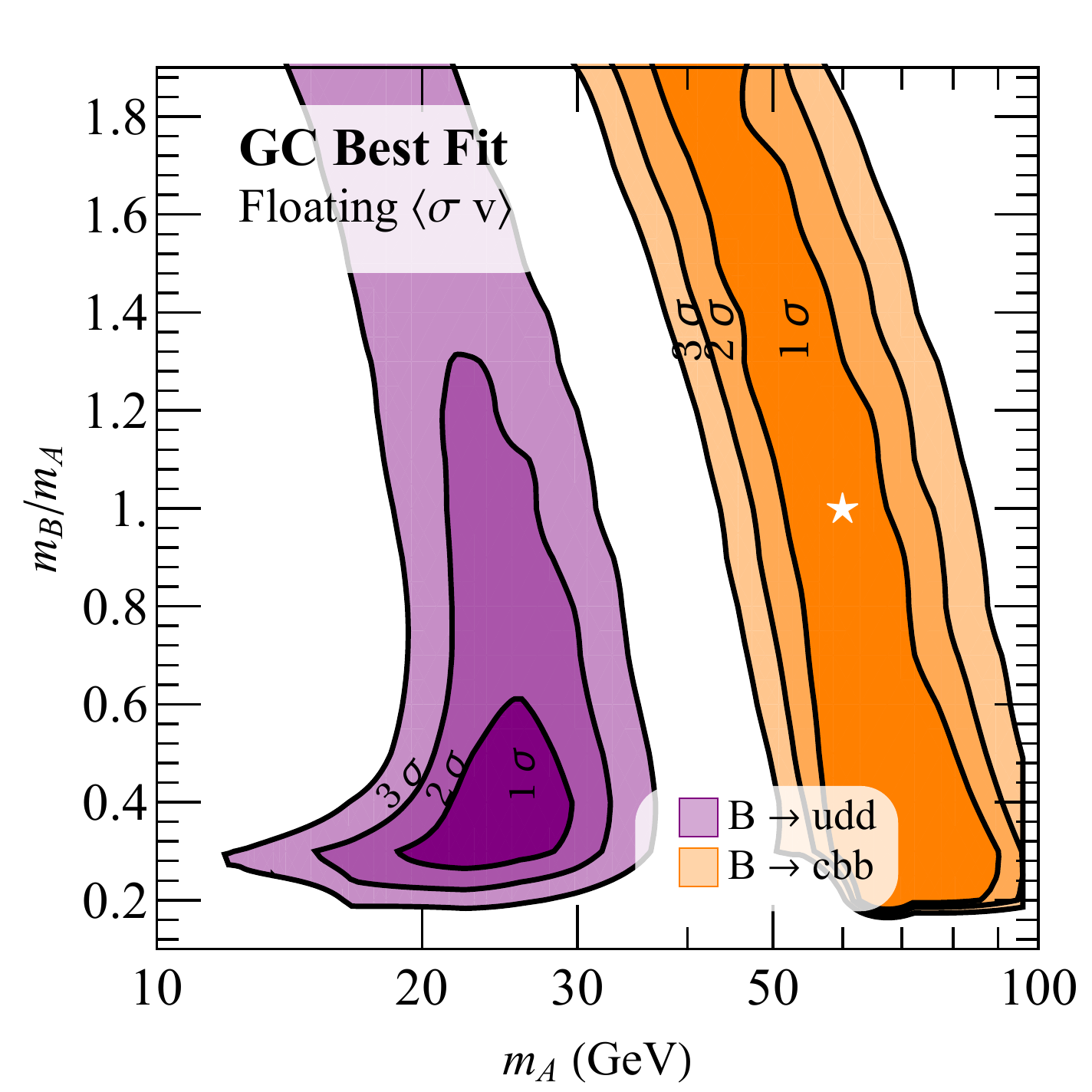} } \label{fig:floatingfits:b}  }
       \caption{Same as \Fig{fig:allplots2}, but floating the mass ratio $m_B/m_A$ in (a) and the effective cross section $\langle \sigma_{AA \ra B \phi} v\rangle_\text{eff}$ in (b).}
   \label{fig:floatingfits}
\end{figure}  

In \Fig{fig:floatingfits}, we show the same parameter space as \Fig{fig:allplots2}, but now letting the third parameter from \Eq{eq:threepara} float to give the best fit.  In \Fig{fig:floatingfits:a}, we note that floating the ratio $m_B/m_A$ extends the best fit to a wider range of values for both $m_A$ and $\langle \sigma v\rangle_\text{eff}$ but still within the vicinity of the benchmark in \Eq{eq:benchmark}.  In \Fig{fig:floatingfits:b}, we see that the best fit for each operator is determined by a band in $m_A$, and the change in the mass ratio and can be compensated somewhat by a change in the cross section.

Finally, returning to \Fig{fig:moneyplotintro} from the introduction, we combine the analysis of the DM and baryon abundances in \Fig{fig:allplots:c} and the GC best fit regions in \Fig{fig:floatingfits:a}.  Note that the abundances are given with respect to the actual cross section $\langle \sigma_{AA \ra B \phi} v \rangle$, while the GC fits and CMB bounds are with respect to the effective cross section $\langle \sigma_{AA \ra B \phi} v \rangle_\text{eff}$.  We made this hybrid choice to avoiding display a pathological region of phase space where one obtains a good fit to the GC excess with a small cross section but overabundant DM.   As advertised, the benchmark parameters in \Eq{eq:benchmark} yield a consistent cosmology and a plausible fit to the GC excess.

\section{Additional Constraints and Signals}
\label{sec:constraints}

Having seen that we can achieve a viable asymmetric DM scenario with intriguing indirect detection signals, we briefly discuss possible additional constraints and signals.

\begin{itemize}
\item \textit{CMB heating bounds}.  The process $AA \to B \phi$ can occur in the early universe, even after thermal freeze-out.  This residual semi-annihilation is constrained by limits on the power injected into the CMB through ionization \cite{Madhavacheril:2013cna}.  The power in this case is parameterized by:
\be
 \label{eq:CMB}
p_\text{CMB} = \frac{f_\text{eff} \langle \sigma_{AA \ra B \phi} v\rangle_\text{eff} }{m_A},
\ee
where $f_\text{eff}$ is the efficiency factor and the effective cross section is defined in \Eq{eq:sigmaeff}.  We can divide $f_{\rm eff}$ into contributions from the $B$ decay products and the (model-dependent) $\phi$ decay products.
\be
f_{\rm eff} = f_{\rm eff}^B + f_{\rm eff}^\phi.
\ee
Following the analysis of \Ref{Slatyer:2015jla}, the efficiency factor from species $X$ is: 
\begin{align}
\label{eq:feffcalc}
f_\text{eff}^X =  \frac{1}{2 m_A} \int_0^{E_X^{\rm max}} E \, \df E & \left[ 2 f_\text{eff}^{e^+ e^-} (E)  \left( \frac{dN}{dE} \right)_{e^+} +f_\text{eff}^{\gamma} (E)  \left( \frac{dN}{dE} \right)_{\gamma}  \right. \nonumber \\ 
&\left. \qquad ~ + f_\text{eff}^{p} (E) \left(  \lp \frac{dN}{dE} \rp_{p} + \lp \frac{dN}{dE} \rp_{\ov{p}}  \right) \right],
\end{align}
where we read off the values of $f_\text{eff}^i$, $i \in \{e^\pm, \gamma\}$ from \Ref{Slatyer:2015jla} and estimate $f_\text{eff}^p \approx  0.2 (f_\text{eff}^{e^+ + e^-} + f_\text{eff}^\gamma)$ following \Ref{Weniger:2013hja}.  For the $B \to cbb$ decay, we find $f^B_\text{eff} = 0.17$.  For the $\phi$ decay, we assume that the dominant decay mode is $\mu^+ \mu^-$, leading to $f^\phi_\text{eff} = 0.07$.

Current constraints from \textit{Planck} \cite{Planck:2015xua} are shown in \Figs{fig:allplots2:a}{fig:floatingfits:a} above (for \Figs{fig:allplots2:b}{fig:floatingfits:b}, the bounds are outside of the plotted region  as they constrain $m_A \lesssim 2$ GeV).  Note that the power injected depends directly on $\langle \sigma_{AA \ra B \phi} v\rangle_\text{eff}$, and therefore directly impacts the GC excess best fit regions.  For a fixed effective cross section, the CMB limits become less stringent as $m_A$ increases.

\item \textit{Antiproton flux bounds}. Though $B$ has baryon number $+1$, we nevertheless expect to obtain antiprotons from the $AA \to B \phi$ process, since the $B$ decay products will hadronize.\footnote{Because of its baryon number, every $B$ decay will necessarily lead to at least one proton (directly or from heavier baryon decay).  Such a proton excess, however, does not seem to be visible over cosmic ray proton backgrounds.}  This additional antiproton flux can be tested in cosmic ray experiments like PAMELA \cite{Adriani:2012paa} and AMS-02 \cite{Aguilar:2015ooa}.  The flux of antiprotons is given by \cite{Cirelli:2010xx}
\be
\frac{d \phi_{\pbar}}{d K} (K, \vec{r}_{\astrosun}) = \frac{v_{\pbar}}{4 \pi} \lp \frac{\rho_{\astrosun}}{m_A} \rp^2 R(K) \, \frac{1}{2} \lang \sigma_{A A \rightarrow B \phi} v \rang_{\rm eff} \frac{\df N_{\pbar}}{\df K}, \label{eq:fluxprotons}
\ee
where $K$ is the kinetic energy of the antiproton (a distinction important in the low energy limit), $v_{\pbar}$ is the velocity of the antiproton, and $R(K)$ is a best fit function that describes the propagation of the antiprotons throughout the galaxy (see \cite{Cirelli:2010xx}).  As in the gamma ray case from \Sec{sec:GC}, we can extract the antiproton spectrum $\df N_{\pbar}/\df K$ from $B$ decays in \pythia and boost to the $AA$ rest frame.

Various groups have derived antiproton bounds using AMS-02 data \cite{Giesen:2015ufa,Jin:2015sqa,Evoli:2015vaa}, typically showing results for WIMPs that annihilate to bottom quarks.  To a reasonable approximation, the antiproton yield in $B \to cbb$ decays is comparable to that of a bottom quark, yielding around 0.3 antiprotons per decay.  More accurately, a single $B \to cbb$ decay from $AA \to B \phi$ yields a factor of 2.5 fewer antiprotons than an equivalent energy $\overline{b} b$ pair.  Thus, instead of evaluating \Eq{eq:fluxprotons} directly, we can simply scale down the $\chi \chi \to \overline{b} b$ bounds by this factor.  Taking the ``Ein MED'' bounds from \cite{Giesen:2015ufa}, for $m_A = 60~\GeV$ we estimate
\be
\label{eq:antiprotonbounds}
\langle \sigma_{AA \ra B \phi} v\rangle_\text{eff} \lesssim 2.3 \times 10^{-26} \text{cm}^3/\text{sec},
\ee
which is in some tension with the GC best fit region.  That said, there is at least a factor of two or three astrophysical uncertainty in these bounds.  In addition, while the CMB bounds are irreducible, in the sense that the same photons that explain the GC excess will inevitably correspond to power injected into the early universe, perturbing the CMB, the antiproton bounds are less directly tied to the gamma ray signal.  For both of these reasons, we have opted not to show the antiproton flux bounds in our plots.

\item \textit{Direct detection bounds}.  Because $A$ has couplings to $\phi$ and $\phi$ couples to SM states, there will necessarily be a contribution to $A$-nucleus scattering from $t$-channel $\phi$ exchange.   For specific models, such bounds might be relevant (and prospects for future direct detection experiments promising), though direct detection constraints can typically be avoided for two reasons.  First, while $\phi$ needs to have large enough coupling to stay in thermal equilibrium with the SM, those required couplings are small from the perspective of $A$-nucleus scattering.  Second, small increases in $m_\phi$ can lead to large decreases in the  $A$-nucleus cross section, since $t$-channel scattering typically scales like $1/m_\phi^4$ at small recoil energies.  

For the specific model studied in \App{app:model}, $\phi$ mixes with the SM Higgs after electroweak symmetry breaking with a mixing angle $\theta_{\phi h}$.  The spin-independent scattering cross section of $A$ through $\phi$ is  (see, e.g.~\cite{Wise:2014jva, Fedderke:2014wda})
\be
\label{eq:directdetectionbound}
\sigma_\text{SI} \simeq \frac{ \lambda_A^2 f^2 m_n^4 \theta_{\phi h}^2}{ \pi \, m_\phi^4 \, v_{\rm EW}^2},
\ee
where $\lambda_A$ is the coupling of $A$ to $\phi$, $v_{\rm EW}$ is the Higgs vacuum expectation value, $f$ is a factor obtained from the different parton fractions which we take to be $0.35$ \cite{Giedt:2009mr}, and $m_n$ is the nucleon mass.  Taking $m_\phi = 250~\MeV$ as a benchmark, this cross section scales like
 \be
 \sigma_{\text{SI}} \simeq 9 \times 10^{-46} \text{ cm}^2 \lp \frac{\lambda_A}{1.0} \rp^2 \lp \frac{250  \text{ MeV}}{m_{\phi}} \rp^4 \lp \frac{\theta_{\phi h}}{  1.1 \times 10^{-7}} \rp^2, 
 \ee
where the baseline value saturates the LUX bound \cite{Akerib:2013tjd}.  While this baseline mixing angle is small, $\phi$ still decays prior to BBN.  Of course, larger mixing angles are allowed for larger $m_\phi$, though one should then account for $\phi$ decays in the GC excess analysis (see footnote \ref{footnote:phimass}).

\item \textit{Flavor bounds}.  One aspect of this scenario that we have not delved into deeply is the generation of the $B$ decay operators in \Eq{eq:decayops}, and in particular their flavor structure.  At the scale $\Lambda$ (or below), some new heavy states are required to generate these operators, and those states could contribute to flavor-violating interactions.  For any $B u^c d^c d^c$-like operator and assuming that $\Lambda$-scale physics conserves CP, one expects that the strongest limits should come from meson mixing induced by those new heavy states \cite{Bona:2007vi,Isidori:2010kg,Kim:2013ivd}.  Since all fields involved in the $B$ decay are right-handed, stronger constraints from chirality-mixing operators can be avoided.  Assuming a generic flavor structure for the new physics, the most constraining bound comes from the $(\bar{c}_R \gamma^\mu u_R)^2 / \Lambda^2$ operator, which is bounded by $\Lambda  \geq 1200~\TeV$ \cite{Isidori:2010kg}.   Taking $m_B = 60~\GeV$ in \Eq{eq:Blifetime}, this yield a $B$ decay width of $\Gamma_B \simeq 1.2 \times 10^{-20}~\GeV$.  Interestingly, $\Gamma_B \sim H$ occurs at a temperature of  $121 ~\MeV$,    above the beginning of BBN at $T \simeq 10~\MeV$ \cite{Wagoner:1966pv,Agashe:2014kda}, so this flavor-safe scenario is indeed  consistent cosmologically.  Smaller $B$ lifetimes (such as for the $\Lambda$ = 300 TeV benchmark we use in \Eq{eq:Blifetime}) require some mild suppression of flavor violation among the heavy states, which is certainly plausible if $\Lambda$-scale physics is approximately flavor conserving.

\item \textit{Collider searches for displaced jets}.    The operators that lead to $B$ decay can also lead to $B$ production at the colliders.  For our $B \to cbb$ benchmark, this cross section is rather small at the LHC, but if the $B \to udd$ operator is present, then there is a more promising process:
\be \label{eq:udBX}
u d \to B \overline{d}.
\ee
For the flavor-safe case of $\Lambda  \simeq 1200~\TeV$, this cross section is negligible, but being optimistic about flavor bounds (and pushing beyond the recommended values in \Fig{fig:Decoupling}), we take $\Lambda \simeq 30~\TeV$ as a benchmark to explore possible LHC signatures.  Since $m_B \simeq 60~\GeV$ and $B$ decays to quarks, this yields a relatively low energy four jet final state, and it is questionable if such events could be seen over overwhelming QCD backgrounds.   That said, plugging $\Lambda = 30~\TeV$ into \Eq{eq:Blifetime}, the $B$ has a lifetime of $\tau_B  \simeq  2.2 \times 10^{-11} \text{ sec}$, or a decay distance of $c \tau_B \simeq 0.7$ cm.  Thus, the jets from $B$ decays come from a (potentially very) displaced vertex, similar to the phenomenology of hidden valleys \cite{Strassler:2006im,Han:2007ae} and their variants (see e.g.~\cite{Freytsis:2014sua,Craig:2015pha,Schwaller:2015gea}). 

To get an estimate of the $B$ production rate, the parton-level cross section in \Eq{eq:udBX} scales like
\begin{eqnarray}
\label{eq:Bxsecpartonlevel}
\sigma (u d \to B \overline{d}) \simeq  \frac{1}{16 \pi}  \frac{( \hat{s} - m_B^2)^2}{4 \hat{s} \Lambda^4} &\simeq& 47 \times 10^{-2} \text{ fb} \lp \frac{\hat{s}}{(14 \text{ TeV})^2} \rp \lp \frac{30  \text{ TeV}}{\Lambda} \rp^4.
\end{eqnarray}
To obtain the proton-proton cross section for the 14 TeV LHC, we integrate over the MSTW2008 LO parton distribution functions \cite {Martin:2009iq}:
\begin{eqnarray}
\sigma (pp \to B + X) &\simeq&  4 \times 10^{-2} \text{ fb} \lp \frac{30  \text{ TeV}}{\Lambda} \rp^4.
\end{eqnarray}
For the high luminosity LHC, with target luminosity of $3~\text{ab}^{-1}$, we expect around 120 events.  With a dedicated displaced jet trigger, one could hope to identify these events.  This would be a distinctive signature for this scenario, especially if one could somehow verify that every $B$ decay yields a baryon $+1$ final state.  We leave a study of the LHC detection prospects to future work (see related studies in \cite{Cui:2014twa,Graham:2012th, Curtin:2015fna}).

\end{itemize}

\section{Conclusions}
\label{sec:conclude}

In this paper, we presented a DM scenario where the interactions responsible for asymmetry sharing in the early universe are potentially visible today through indirect detection experiments.  The key novelty compared to other asymmetric DM scenarios is that the DM abundance is set by thermal freeze-out involving an unstable particle $B$ rather than by the decoupling of high-scale interactions.   Assuming that $B$ chemically decouples above the DM mass, the parametrics of $A$ freeze-out behaves much like a standard WIMP, albeit with a non-zero chemical potential and a rescaled freeze-out value $x_F \to y_F$.  Given this connection to WIMP physics, it is not surprising that we found a benchmark scenario with the right DM and baryon abundances.  Intriguingly, this same benchmark yields a gamma spectrum from $AA \to B \phi$ semi-annihilation compatible with the GC gamma ray excess seen by \textit{Fermi}.

This work emphasizes that non-minimal asymmetric DM scenarios can produce interesting indirect detection signals without relying on a residual symmetric DM component.  As shown already in \Refs{Pearce:2013ola,Detmold:2014qqa,Pearce:2015zca}, asymmetric DM models with multiple stable states can yield indirect detection signals from semi-annihilation.  Here, we emphasize the role that unstable dark sector states can play both in generating indirect detection signals as well as in sharing primordial asymmetries.
 
An interesting variant to our scenario is if $B$ were stable, perhaps due to an additional $\mathbf{Z}_2$ symmetry.  In that case, both $A$ and $B$ would contribute to the DM abundance, with the relative ratio determined by the $AA \to B \phi$ process.  As long as there is a sufficient abundance of $A$ particles, then this semi-annihilation process could give rise to both a boosted DM signal from the final state $B$ \cite{Agashe:2014yua,Berger:2014sqa,Kong:2014mia,Kopp:2015bfa} as well as a standard indirect detection signal from the decays of $\phi$.  We leave a study of this scenario to future work.

Finally, the nature of DM cannot be determined through any single observation, and even if the GC excess is indeed due to DM (semi-)annihilation, one would want to determine the dark sector properties through a suite of other experiments.  For this particular DM scenario, the hadronic $B$ decays imply an irreducible cosmic ray signal from antiprotons and potentially antideuterons.  More intriguingly, evidence for $B$ particles might show up at the LHC or future colliders through displaced hadronic decays, yielding a visible portal to a rich dark sector.

\begin{acknowledgments}
We thank Matthew McCullough for collaborating in the initial stages of this work.  We benefitted from discussions with George Brova, Gilly Elor, Alexander Ji, Andrew Larkoski, Ian Moult, Marieke Postma, Nicholas Rodd, Tracy Slatyer, Iain Stewart, Wei Xue, and Kathryn Zurek.  This work was supported by the U.S. Department of Energy (DOE) under cooperative research agreement DE-SC-00012567. N.F.\ is also supported by Funda\c{c}\~{a}o de Amparo \`{a} Pesquisa do Estado de S\~{a}o Paulo (FAPESP) and Conselho Nacional de Ci\^encia e Tecnologia (CNPq).  J.T. is also supported by the DOE Early Career research program DE-SC-0006389 and by a Sloan Research Fellowship from the Alfred P. Sloan Foundation.
 
\end{acknowledgments}

\appendix
\section{Concrete Model}
\label{app:model}

\allowdisplaybreaks[1]

 \begin{table}[t]
\begin{center}
\begin{tabular}{cccc}
\hline
\hline
Field & Spin & Baryon number \\ 
\hline
$\psi_A$ & Weyl left & $+1/2$   \\
$\psi_A^c$ & Weyl left & $-1/2$   \\
$\phi_{A}$ & complex scalar  & $+1/2$   \\
$\phi_{A}^c$ & complex scalar  & $-1/2$   \\
$B$ & Weyl left  & $+1$  \\
$B^c$ & Weyl left  & $-1$  \\
$\phi$ & complex scalar & $0$ \\ 
\hline
\hline
\end{tabular}
\end{center}
\caption{Field content for the concrete model, which generate the required interactions specified in \Figs{fig:interactions}{fig:interactionsBSM}.}
\label{tab:fieldcontent}
\end{table}

For simplicity in the text, we referred to $A$ as being a single particle.  Since the decay operators in \Eq{eq:decayops} imply that $B$ must be fermion, though, $A$ should be replaced by a fermion/boson system of $\psi_A/\phi_A$.  Motivated partly by supersymmetry but mainly by the resulting simplification of the Boltzmann equations, we take $\psi_A/\psi_A^c$ to be a Dirac fermion that is mass degenerate along with two complex scalars $\phi_{A}/\phi_{A}^c$ (i.e.\ two chiral multiplets with a holomorphic mass).  We comment on the impact of splitting the $\psi_A/\phi_A$ masses below.

In the text, we also referred to $\phi$ as being a single particle and used the benchmark $m_\phi = 250~\MeV$ for studying CMB and direct detection bounds.  Here we replace $\phi$ with two real fields, one scalar and one pseudoscalar.  This can also be motivated by supersymmetry, but more relevant to our scenario, if $\phi$ were composed of a single real scalar field, then the annihilation $\psi_A \overline{\psi_A} \to \phi \phi$ would be $p$-wave suppressed.  We consider these two fields to have somewhat heavier masses than considered in the text, in order to avoid additional bound state formation (see footnote \ref{footnote:bound}).  We will see that this has a negligible effect on the CMB bounds  in \Eq{eq:feffcalc} and weakens slightly the direct detection bounds in \Eq{eq:directdetectionbound}.
 
The field content for this model is shown in \Tab{tab:fieldcontent}, where $B/B^c$ is a Dirac fermion and $\phi$ is a complex scalar.  We separate $\phi$ into its scalar component $s$ and pseudoscalar component $a$, as
\be
\phi = \frac{s + i a}{\sqrt{2}}.
\ee
This is similar to the axion portal \cite{Nomura:2008ru,Mardon:2009gw}, though we have suppressed a possible vacuum expectation value (vev) for $\phi$ in order to remain agnostic as to whether or not $a$ is a pseudo-Goldstone boson from spontaneous symmetry breaking.  The Lagrangian is then given by
\begin{align}
 \mathcal{L}_{\text{free}} &= i\overline{\psi}_A \overline{\sigma}^\mu \partial_\mu \psi_A +  i \overline{\psi}_A^c \overline{\sigma}^\mu \partial_\mu \psi_A^c   - \, (m_A \psi_A^c \psi_A + \text{h.c.}) \nonumber \\ 
  &\quad ~ + | \partial_\mu \phi_{A} |^2 + | \partial_\mu \phi_{A}^c |^2 - m_A^2 (|\phi_{A}|^2 + |\phi^c_{A}|^2) \nonumber \\
   &\quad ~ +   i\overline{B} \overline{\sigma}^\mu \partial_\mu B + i \overline{B}^c \overline{\sigma}^\mu \partial_\mu B^c - (m_B B B^c + \text{h.c.}) \nonumber \\
  &\quad ~ + |\partial_\mu \phi|^2 - \frac{1}{2} m_{s}^2 s^2 - \frac{1}{2} m_a^2 a^2, \\ \nonumber
\mathcal{L}_{\text{int}}^4 &=  \lambda_{AB} ( \psi_A B^c \phi_A + \psi_A^c B \phi_{A}^c +  \text{h.c.}) \\
 &\quad  ~ - (\lambda_A \psi_A \psi_A^c  \phi + \lambda_B B B^c  \phi + \text{h.c.}) \nonumber \\
 &\quad ~ + \mu_{A \phi} \left( |\phi_{A}|^2 + |\phi_{A}^c|^2 \right) (\phi + \phi^\dagger)  + \lambda_{A \phi}  \left( |\phi_{A}|^2 + |\phi_{A}^c|^2 \right) \phi^\dagger \phi \nonumber \\
 &\quad ~ - V(H,\phi), \label{eq:L4int} \\ 
  \mathcal{L}_{\text{int}}^6 &= \frac{B u^c d^c d^c}{\Lambda^2} + \text{h.c.} \,. \label{eq:L6int}
 \end{align}
Note that we have introduced a mass splitting between the scalar and pseudoscalar modes, such that the decay $s \to aa$ is kinematically allowed.  For simplicity, we have taken the Yukawa couplings to $s$ and $a$ to be the same, and assumed various relations among the scalar $\phi_A$ and $\phi_A^c$ couplings.
 
The potential term $V(H,\phi)$ includes terms that mix $\phi$ with the Higgs sector after electroweak symmetry breaking.  Motivated by supersymmetry and the axion portal, we assume a two Higgs doublet model with $H_{u,d}$, such that one possible mixing term is
\be \label{eq:muterm}
V(H,\phi) \supset \mu_{\phi H} \, \phi \, H_u H_d + \text{h.c.}   
\ee
This interaction assures that both $s$ and $a$ can decay to SM states, though $V(H,\phi)$ generically contains cubic interactions such that the $s \to aa$ decay mode dominates.  We ignore the induced vev of $\phi$ from this mixing, since it can be absorbed into a redefinition of the other couplings.  Depending on the potential, the Higgs vevs can also contribute to the $\phi$ mass, though we expect this to be a small effect given the small mixing angle suggested by \Eq{eq:directdetectionbound}.  Without supersymmetry, some degree of fine tuning would be necessary to keep the $s$ and $a$ masses small given their large couplings to the $A$ and $B$ states.

The baryon assignment of the different particles is set by $B u^c d^c d^c/\Lambda^2$ in \Eq{eq:L6int}.  So unlike in the SM, baryon number is not an accidental symmetry of this Lagrangian, though that conclusion might change depending on the precise dynamics present at $\Lambda$.

For the purposes of \App{app:Boltzmann}, the key feature of the coupling choices above is that the $A$ scalars are treated symmetrically, such that any process involving $\phi_A$ has a counterpart involving $(\phi_A^c)^\dagger$.  This, along with the assumed mass degeneracy of the $A$ fermions and $A$ scalars, allows the full Boltzmann system to simplify to a two-particle system.  We checked that all the relevant DM interactions from this Lagrangian include an $s$-wave term (see e.g.~\cite{Cui:2010ud,Kumar:2013iva}), which is necessary to achieve the desired cosmology.  For example, an $s$-wave annihilation channel for the $A$ fermions is possible via $\psi_A \overline{\psi_A} \to s a$.

Introducing mass (or coupling) splittings between $\psi_A$ and $\phi_A$ would add new parameters to the model studied and complicate the Boltzmann equations.  The dominant effect of such a splitting is to change the resulting populations of the $\psi_A$ and $\phi_A$ components.  To maintain roughly the same phenomenology as presented in the text, we need $m_{\psi_A}/m_{\phi_A}$ to be $\mathcal{O}(1)$, such that the lighter DM component is still Boltzmann suppressed at freezeout.  We also need to satisfy the mass ordering
\be
\max\{m_{\psi_A},m_{\phi_A}\} \lesssim m_B < m_{\psi_A} + m_{\phi_A}.
\ee
This first inequality comes from the requirement of not depleting $A$ particles by scattering with $\Bbar$ particles prior to freezeout (see footnote \ref{footnote:splitmassBbar}), and the second inequality ensures that the semi-annihilation process $\psi_A \phi_A \rightarrow B \phi$ is kinematically allowed even at threshold.  

In addition to the possibility that $B$ might correspond to a bound state of $A$ particles, the $\phi$ mediator might give rise to additional bound states of $A$ and/or $B$.  For simplicity of our analysis, we want to avoid the possibility of $\phi$-mediated bound states, which would further complicate the Boltzmann analysis.  Note that only the scalar $s$ mediates a $1/r$ Yukawa potential between the $A$ particles, whereas the pseudoscalar $a$ can only mediate a $1/r^3$ spin-dependent potential among the $A$ fermions.  For our benchmark, with $\mathcal{O}(1)$ $\phi$ couplings and DM masses around $50~\GeV$, we can avoid bound state formation if $m_s \gtrsim \mathcal{O}(1~\GeV)$.

The presence of both $s$ and $a$ turns out to have relatively little effect on the phenomenology presented in the body of the paper.  There are now two semi-annihilation processes relevant for indirect detection, $\psi_A \phi_A \rightarrow B a$ and $\psi_A \phi_A \rightarrow B s$.  As long as $2 m_\mu < m_a < 3 m_\pi$ (such that $a$ dominantly decays as $a \to \mu^+ \mu^-$) and $m_s > 2 m_a$ (such that $s$ dominantly decays  as $s \to aa \to \mu^+ \mu^- \mu^+ \mu^-$), then the resulting photons from muon FSR give subdominant contributions to the photon flux as needed for \Sec{sec:GC}.  The CMB bounds in \Eq{eq:feffcalc} only have a mild dependence on the injected muon spectrum so are largely unchanged.  Raising $m_s$ weakens the spin-independent direct detection bound in \Eq{eq:directdetectionbound} and spin-dependent bounds from $a$ exchange are typically subdominant.

Finally, we note that similar phenomenology could be achieved by replacing $\phi$ with a $\text{U}(1)'$ gauge field that kinetically mixes with SM hypercharge \cite{Holdom:1985ag,Okun:1982xi,Galison:1983pa}.  The main challenge for using this hypercharge portal is that the $B$ decay operator in \Eq{eq:L6int} would then require an insertion of the $\text{U}(1)'$-breaking Higgs field, making it an even higher dimension operator.

\section{Chemical Potential Analysis}
\label{app:chemical}

To determine the abundance of DM at the decoupling temperature $T_D$, we have to consider the chemical potentials of all relevant SM and dark species.  For simplicity, we work in a regime where the sphaleron process becomes inactive prior to $T_D$, such that we can treat the baryon and lepton asymmetries as being independently conserved during $B$ decoupling.  The states in the dark sector then carry effective baryon number consistent with the operators in \Eq{eq:decayops}.

In the early universe, SM interactions guarantee chemical equilibrium among SM particles.  After electroweak symmetry breaking, the chemical potential relations can be written as \cite{Harvey:1990qw,Ibe:2011hq} 
\begin{align}
\mu_u - \mu_{d} &=  \mu_\nu - \mu_{e} = \mu_W, \label{eq:chemicalSM1} \\ 
-3(\mu_{u} + \mu_d) & = \mu_\nu + \mu_{e}  \label{eq:chemicalSM2},
\end{align}
where $\mu_u$, $\mu_d$, $\mu_\nu$, and $\mu_e$ refer to the chemical potentials of each flavor of up-type quark ($u_L$ and $u_R$), down-type quark ($d_L$ and $d_R$), left-handed neutrino ($\nu_L$), and charged lepton ($e_L$ and $e_R$).  Note that \Eq{eq:chemicalSM1} is enforced by $W^{\pm}$ exchanges (with chemical potential $\mu_W$) and \Eq{eq:chemicalSM2} is imposed by the sphaleron process.  At temperatures below the top quark mass, the neutrality condition imposes
 \begin{equation} \label{eq:chemicalQ}
8 \mu_u - 6 \mu_d - 6 \mu_e + 6 \mu_W = 0.
\end{equation}
The above relations imply
\be
\label{eq:chemqfinal}
\mu_u = \frac{3}{13}\mu_d = -\frac{3}{19} \mu_e,
\ee
which allows us to write the chemical potential for SM baryons as
\begin{equation}
\label{eq:baryonnumber}
\mu_b \equiv (3-1) (\mu_{u_L} + \mu_{u_R})+ 3 (\mu_{d_L}+ \mu_{d_R})= -\frac{90}{19} \mu_e\,,
\end{equation}
where we do not include the contribution from the top quark.  The semi-annihilation process in \Eq{eq:semiannihilation} and the transfer operators in \Eq{eq:decayops} impose the relations
\begin{align} 
\mu_B & = 2 \mu_A, \\ \label{eq:eq22}
\mu_B &= \mu_u + 2 \mu_d.
\end{align}
Using \Eqs{eq:chemqfinal}{eq:baryonnumber}, the chemical potentials for $A$ and $B$ can be simplified to
\be
\label{eq:chemDMfinal}
\mu_B = 2 \mu_A = \frac{29}{90} \mu_b .
\ee

For a species $i$ with $g_i$ degrees of freedom and mass $m_i$, the relation between the charge density and the chemical potential $\mu_i$ at temperature $T$ is \cite{Barr:1990ca} (see also \cite{Servant:2013uwa})  
\begin{equation} \label{eq:ndensity}
 n_{i} - n_{\overline{i}}  = g_i \, f(m_i/T) \, T^3 \left(\frac{\mu_i}{T}\right)\,,
\end{equation}
where we are assuming $\mu_i\ll T$ with
\begin{equation} \label{eq:fx}
f(x) =  \left\{ \begin{array}{l}
\frac{1}{4 \pi^2} \int_{m_i/T}^\infty \frac{y^2 dy}{\cosh^2{(\frac{1}{2}\sqrt{x^2 + y^2})}} \equiv f_{\text{f}}(x) \quad \text{(for  fermions)}, \\
\frac{1}{4 \pi^2} \int_{m_i/T}^\infty \frac{y^2 dy}{\sinh^2{(\frac{1}{2}\sqrt{x^2 + y^2})}}  \equiv f_{\text{b}}(x) \quad \text{(for bosons)}.
\end{array}  \right.
\end{equation}
The function $f(x)$ takes into account the Boltzmann suppression of particle $i$. In the limit $x \gg 1$, $f(x)   \simeq 2 (x/2 \pi)^{3/2} e^{-x} $.  Using  Eqs.~(\ref{eq:baryonnumber})--(\ref{eq:fx}), we can write the asymmetries as $\eta_i = X_i(n_i - n_{\overline{i}})/s$, where $X_i$ denotes the baryon charge of species $i$.  This leads to   
\begin{eqnarray} \label{eq:etabaryon}
\eta_{b} &=& \frac{1}{3}\frac{g_q \, T^2 \mu_b}{s}  f_{\rm f}(0)\,, \\ \label{eq:etapsiA}
\eta_{\psi_A}&=&  \frac{1}{2}\frac{29}{180} \frac{ g_{\psi_A} \, T^2 \mu_b}{s}  f_{\rm f}(m_A/T) \,,\\
\label{eq:etaphiA}
\eta_{\phi^{(c)}_{A}}  &=&  \frac{1}{2} \frac{29}{180} \frac{ g_{\phi^{(c)}_{A}} \, T^2 \mu_b}{s}  f_{\rm b}(m_A/T)  \,,\\
\label{eq:etaB}
\eta_{B}  &=& \frac{29}{90} \frac{g_B \, T^2 \mu_b}{s}  f_{\rm f}(m_B/T)\,,
\end{eqnarray}
where $g_q$ is the number of degrees of freedom for each quark flavor, i.e.\ $g_q = 3$ from color (since the left- and right-helicities are already included in \Eq{eq:baryonnumber}).  For the system described in \App{app:model},
\be
g_{\psi_A} = 2, \qquad g_{\phi_{A}} = 1, \qquad g_{\phi^c_{A}} = 1, \qquad g_B = 2.
\ee

The relations above are valid for temperatures $T \geq T_D$, where $T_D$ is the $B$ decoupling temperature.  Using
\be
\eta_{A} \equiv \eta_{\psi_A} + \eta_{\phi_{A}} + \eta_{\phi_{A}^c},
\ee
the total asymmetry $\eta_{\text{tot}}$ in \Eq{eq:asymmetries} can be written as
\begin{eqnarray} \label{eq:etatot}
\eta_{\rm tot} &=&   \eta_{A} + \eta_{B} + \eta_{b} \nonumber \\
 &=&\eta_b \left[ \frac{29}{360} \frac{ 3\, \left(g_{\psi_A} \,f_{\rm f}(m_A/T)+ 2 \, g_{\phi_A} \,f_{\rm b}(m_A/T)\right)}{g_q\,  f_{\rm f}(0)} + \frac{29}{90} \frac{3 \,g_B \,f_{\rm f}(m_B/T)}{g_q \,  f_{\rm f}(0)} + 1  \right] , \nonumber \\
\end{eqnarray}
which is the basis for Eqs.~(\ref{eq:etaA})--(\ref{eq:etab}). 
From  \Eq{eq:ndensity} and the relation above, it is also possible to write the DM chemical potentials as functions of $\eta_{\rm tot}$ as
 \be
 \frac{\mu_i}{T} =    \left(\frac{2 \pi^2 g_{*s}}{ 45 g_i  }\right) \frac{\eta_i}{f(m_i/T)}, \label{eq:chempot}
 \ee
where we used $s= 2\pi^2 g_{*s} T^3/45$.

\section{Thermal Cross Section for $B \overline{q} \rightarrow qq$}
\label{app:crosssection}

The process $B \overline{q} \rightarrow qq$, where $q$ is a quark consistent with the operators in \Eq{eq:decayops}, is efficient at high temperatures, and must be included when determining the $B$ decoupling temperature $T_D$.  From the amplitude
\be
i \mathcal{M} = \frac{-i}{\Lambda^2} (x_1 x_2) (y_3 y_4)
\ee
expressed in terms of the Weyl wavefunctions $x_i$ and $y_i$, we find
\be
\frac{1}{4} \sum |\mathcal{M}|^2 = \frac{s(s-m_B^2)}{4 \Lambda^4},
\ee
where $s$ is the square of the center-of-mass energy, and we assume $m_B\gg m_q$. The differential cross section with respect to the Mandelstam $t$ variable is
\be
\frac{d\sigma}{dt} = \frac{1}{16 \pi} \frac{|\mathcal{M}|^2}{ \lambda (s,m_B^2,0)},
\ee
where to simplify the phase space factor we use the function
\be
\lambda(x,y,z) = x^2 + y^2 + z^2 - 2 xy - 2xz - 2 yz.
\ee
Integrating the cross section over $t$, we find
\be
\sigma = \frac{s}{64 \pi \Lambda^4}.
\ee

At sufficiently high temperatures, we can ignore the chemical potentials of $B$ and the quarks.  For a generic scattering process $12 \to 34$, the M\o ller thermal cross section is \cite{Gondolo:1990dk}
\be 
\langle \sigma v_\text{Mol} \rangle_{12 \rightarrow 34} = \frac{1}{n_{1}^\eq n_\text{2}^\eq} \int \frac{g_1 \, \df^3 p_1}{(2 \pi)^3} \frac{g_2 \, \df^3 p_2}{(2 \pi)^3} \sigma v_{\text{Mol}} e^{-(E_1 + E_2) /T},  \label{eq:sigma22}
\ee
where the M\o ller velocity is defined such that the product $v_{\text{Mol}}\, n_1 n_2$ is Lorentz invariant, and 
\be
n_{i}^\eq(T) = \int \frac{g_i \, \df^3 p_i}{(2 \pi)^3}  e^{-E /T} \,,
\ee
In our case, $n_1^\eq = n_{B0}^\eq$ is the thermal Boltzmann distribution for $B$,
\begin{equation} \label{eq:nB0}
n_{B0}^\eq = \frac{g_B}{2 \pi^2} \left(\frac{m_B}{T}\right)^2 K_2\left(\frac{m_B}{T}\right) T^3,
\end{equation}
where $K_i$ is the modified Bessel function of order $i$, and $n_2 = n_{q0}^\eq$ is the relativistic distribution for quarks,
\be
n_{q0} = \frac{3}{4} \frac{\zeta(3)}{\pi^2} g_q T^3, \label{eq:SMeq}
\ee
where here we have corrected for the Fermi statistics of the relativistic quarks. The relative velocity $v_\text{Mol}$ is obtained from $p_1 \cdot p_2 = v_\text{Mol} E_1 E_2$.  The integral in \Eq{eq:sigma22} can be reparameterized following the annihilation example of \Ref{Gondolo:1990dk} as
\be
\df^3 p_1 \, \df^3p_2 = 2 \pi^2 E_1 E_2  \, \df E_{+} \,  \df E_{-} \,  \df s,
\ee
with the limits of integration:
\begin{eqnarray} \label{eq:EpLimit}
E_+  &=& E_1 + E_2 \geq \sqrt{s}, \\  \label{eq:EmLimit}
E_-  &=& E_1 - E_2 ~~\text{set by the condition } |\cos \theta| \leq 1, \\  \label{eq:sLimit}
s &\geq& m_B^2. 
\end{eqnarray}
Performing these integrals, $\langle \sigma v_\text{Mol} \rangle_{B \overline{q} \rightarrow qq}$ is given by
\begin{eqnarray}
\langle \sigma v_\text{Mol}\rangle_{B \overline{q} \rightarrow qq} &=& \frac{1}{n_{1}^\eq n_\text{2}^\eq} \frac{g_B g_q}{(2 \pi)^6}\frac{\pi T}{32 \, \Lambda^4}\int_{m_B^2}^\infty \df s \sqrt{s}(s - m_B^2)^2 \, K_1\left(\frac{\sqrt{s}}{T}\right) \nonumber 
\\ \label{eq:sigmaf}
&=& \frac{m_B^2 K_4\left(\frac{m_B}{T}\right)}{48  \pi \Lambda^4 \zeta(3) K_2\left(\frac{m_B}{T}\right)} \,.
\end{eqnarray}

\section{Boltzmann Equation Details}
\label{app:Boltzmann} 

\allowdisplaybreaks[1]

In going from the full model in \App{app:model} to the simplified discussion in \Sec{sec:model}, we asserted that the four-particle system $X \in \{\psi_A, \phi_{A}, \phi_{A}^c, B\}$ could be reduced to an effective two-particle system.  Indeed, this is possible if all states type $A$  are mass degenerate and have equal chemical potentials such that $\mu_{\psi_A}= \mu_{\phi_{A}}=\mu_{\phi_{A}^{c\da}}\equiv \mu_{A}$.  We now derive the full Boltzmann system and show how this reduction occurs.

Unlike the discussion in \Sec{subsec:Afreezeout} where we introduced the $y$ variable, here we stick to the notation $x = m_A/T$.  The equilibrium distributions for a state $X$ are given by
\be \label{eq:equilibrium}
Y_X^\eq (x) = Y_{X0}^\eq (x) \exp(\mu_X/T),
\ee
where $\mu_X$ is the chemical potential of $X$ and we assume $\mu_X \ll T$ which is a valid assumption until freezeout.  In the non-relativistic limit ($m_X \gg T$), bosons and fermions follow the same distribution \cite{Kolb:1990vq}, and the equilibrium functions denoted with a subscript ``0" are
\be \label{eq:equilibrium0}
Y_{X0}^{\rm{eq}} = \frac{g_X}{g_{*s}} \frac{45}{4 \pi^4}  x^2 K_2 \left(x \right).
\ee
We use the notation $\lambda = s/H(m_A)$, where $s$ is the entropy of the universe, $H$ the Hubble constant at the temperature $T =  m_A$, and $g_{*s}$ is the effective number of relativistic degrees of freedom at the same temperature \cite{Kolb:1990vq}.   The chemical potentials satisfy $2 \mu_A = \mu_B$ when the semi-annihilation process $AA \ra B \phi$ is in equilibrium.  We also have $\mu_A = -\mu_{\Abar}$ and $\mu_B = -\mu_{\Bbar}$ when the annihilation processes $A\Abar \ra \phi \phi$ and $B\Bbar \ra \phi \phi$ are in equilibrium.  Here we assume $\mu_\phi = 0$, since the $\phi$ is in thermal equilibrium with the SM and there is no symmetry forbidding interactions like $\phi \phi \ra \phi \phi \phi$. 

By assumption in \App{app:model}, the couplings involving ${\phi_{A}}$ and $\phi_{A}^{c\da}$ are identical, so we know that
\be
Y_{\phi_{A}} = Y_{\phi^{c\da}_{A}}
\ee
at all temperatures $T$.  Therefore, processes that involve $\phi_{A}$, such as $\langle \sigma_{ \ov{\psi_A} B \ra \phi_{A} \phi  } v \rangle $, will simply get a factor of 2 from including the equivalent process with $\phi_{A}^{c\da}$.  With this simplification, the Boltzmann equations for the $\{\psi_A, B \}$ system are:
\begin{align}
 \frac{\df Y_{\psi_A}}{\df x} = -\frac{\lambda}{x^2} \Biggl[ &\langle \sigma_{\psi_A \overline{\psi_A} \ra \phi \phi} v  \rangle \Bigl(Y_{\psi_A} Y_{\ov{\psi_A}} - Y_{\psi_A}^{\rm{eq}} Y_{\ov{\psi_A}}^\text{eq}  \Bigr) \nonumber \\
 \qquad  \qquad ~ + 2\, & \langle \sigma_{\psi_A \phi_{A} \ra B \phi} v  \rangle \Biggl(Y_{\psi_A} Y_{\phi_{A}} - Y_{B} \frac{\, Y_{\psi_A}^{\rm{eq}} \,  Y_{\phi_{A}}^\eq }{Y_{B}^{\rm{eq}}  } \Biggr) \label{eq:BoltA} \nonumber \\ 
\qquad  \qquad~ - 2\, & \langle \sigma_{\phi_{A}^\da B \ra \psi_A \phi} v   \rangle \Biggl(Y_{\phi_{A}^\da} Y_B - Y_{\psi_A}  \frac{Y_B^\eq \,   Y_{\phi_{A}^\da}^\eq  }{ Y_{\psi_A}^\eq}    \Biggr)
 \nonumber \\
 \qquad  \qquad~ + 2\, & \langle \sigma_{ \psi_A \ov{B} \ra \phi_{A}^\da \phi  } v   \rangle \Biggl(Y_{\psi_A} Y_{\ov{B}}   - Y_{\phi_{A}^\da }     \frac{Y_{\psi_A}^\eq Y_{\ov{B}}^\eq}{Y_{\phi_{A}^\da}^\eq}       \Biggr)      \Biggr] , \\ 
\frac{\df Y_B}{\df x} = -\frac{\lambda}{x^2} \Biggl[ &\langle \sigma_{B \ov{B} \ra \phi\phi} v  \rangle (Y_B Y_{\ov{B}} - Y_{B}^\eq Y_{\Bbar}^\eq)    \nonumber \\
 \qquad  \qquad~ - 2 \,& \langle \sigma_{\psi_A \phi_{A} \ra B \phi} v    \rangle \Biggl(Y_{\psi_A} Y_{\phi_{A}}  - Y_{B} \frac{Y_{\psi_A}^{\rm{eq}}\,  Y_{\phi_{A}}^\eq }{Y_{B}^{\rm{eq}}  } \Biggr)
\nonumber \\
 \qquad  \qquad~ + 2 \, & \langle \sigma_{ \ov{\psi_A} B \ra \phi_{A} \phi  } v   \rangle \Biggl(Y_{\ov{\psi_A}} Y_{B}   - Y_{\phi_{A} }    \frac{Y_{\ov{\psi_A}}^\eq Y_{B}^\eq}{Y_{\phi_{A}^c}^\eq}  \Biggr)\nonumber\\
 \qquad  \qquad~ +  2 \,  & \langle \sigma_{ \phi_A^\da B \ra \psi_{A} \phi  } v  \rangle \Biggl(Y_{\phi_{A}^\da}  Y_{B}   -  Y_{\psi_{A} }    \frac{Y_{\phi_A^\da}^\eq Y_{B}^\eq}{Y_{\psi_{A}}^\eq}  \Biggr)   \nonumber \\ 
 \qquad  \qquad~  +\, & \langle \sigma_{B \overline{q} \rightarrow qq} v  \rangle  Y_\text{SM}^\eq \lp Y_B - Y_B^\eq \rp   \Biggr]  - \frac{\Gamma_B x }{H} \lp Y_B - Y_B^{\text{eq}}\rp  \label{eq:BoltB} .  
\end{align}
The Boltzmann equations for $\phi_{A}$ can be easily extrapolated from those of $\psi_A$.  Here, we have neglected the conversion process $A \Abar \rightarrow B \Bbar$ since annihilation dominates over conversions for determining the freezeout dynamics of near-mass states (recall that our mass range of interest is $m_A \lesssim m_B < 2 m_A$).

At high enough temperatures (though not so high that boson/fermion statistics matter), the equilibrium distributions satisfy
\be
Y_{\psi_A}^\eq = 2 \, Y_{\phi_{A}}^\eq,
\ee
where the factor of 2 comes from the number of degrees of freedom.  In general, though, the evolution of $Y_{\psi_A}$ and $2  \, Y_{\phi_{A}}$ will not remain identical, except for special choices of the couplings in \Eq{eq:L4int}.  To see which combination is needed, consider the evolution of $Y_{\psi_A} - Y_{\phi_{A}} - Y_{\phi_{A}^{c\da}} = Y_{\psi_A} - 2 Y_{\phi_{A}}$.
\begin{align}
          \frac{\df}{\df x} \lp Y_{\psi_A} - 2 Y_{\phi_{A}} \rp &= -\frac{\lambda}{x^2} \Biggl[ \langle \sigma_{\psi_A \overline{\psi_A} \ra \phi \phi} v \rangle (Y_{\psi_A} Y_{\ov{\psi_A}}) - 2 \, \langle \sigma_{\phi_{A} \phi_{A}^\da \ra \phi \phi } v \rangle (Y_{\phi_{A}} Y_{\phi_{A}^\dagger }  ) \nonumber \\ 
 & \hspace{-1.5cm} ~+ 2 \,  \langle \sigma_{\psi_A \overline{B} \ra \phi_A^\da \phi} v \rangle \Biggl( (Y_{\psi_A} - 2 \, Y_{\phi_{A}} ) Y_{\overline{B}}  - \lp Y_{\phi_{A}}^\da - 2 \times \frac{1}{4} \, Y_{\overline{\psi}_A} \rp  \frac{ Y_{\psi_A}^\eq Y_{\overline{B}}^\eq }{ Y_{\phi_A^{\da}}^\eq}    \Biggr)  \nonumber \\ 
 & \hspace{-1.5cm} ~ - 2 \, \langle \sigma_{\phi_{A}^\da B \ra \psi_A \phi} v \rangle \Biggl( \lp Y_{\phi_{A}^\da} - \frac{1}{2} \, Y_{\overline{\psi}_A} \rp Y_{B} - \lp Y_{\psi_A} - \frac{1}{2} \times 4 \, Y_{\phi_{A}}  \rp   \frac{Y_{\phi_A^\da}^\eq Y_{B}^\eq }{ Y_{\psi_A}^\eq}    \Biggr)    \Biggr],  \label{eq:boltDif}
\end{align}
In addition to the factors of 2 coming from including ${\phi_{A}} \leftrightarrow \phi_{A}^{c\da}$, there is an additional factor of 2 in the second line coming from averaging over polarizations in the cross section, since
\be
\langle \sigma_{\phi_{A} \overline{B} \ra \ov{\psi_A} \phi}  v \rangle  = 2 \, \langle \sigma_{\psi_A \overline{B} \ra \phi_{A}^\da \phi} v \rangle.
\ee
There is a factor of $1/4$ in the second line from the ratios of the equilibrium distributions since $Y_{\psi_A}^\eq = 2 Y_{\phi_{A}}^\eq$.   In the third line, the factor of $1/2$ comes from averaging over spins in the cross sections, and the factor of 4 comes from the ratio of equilibrium densities.  From the semi-annihilation terms alone (i.e.\ second and third lines), it is now clear that if $Y_{\psi_A}^\eq = 2 Y_{\phi_{A}}^\eq$ at early times, then this relation will continue to hold at later times.   For the annihilation terms (i.e.\ first line) to maintain the same relation, we simply need to choose couplings such that
\be
\label{eq:neededcouplings}
\langle \sigma_{\phi_A \phi_A^\da \ra \phi \phi} v \rangle  = 2 \, \langle \sigma_{\psi_A \ov{\psi_A} \ra \phi \phi} v\rangle.
\ee
This in turn sets the needed couplings in \Eq{eq:L4int}.  Of course, for more general couplings, one can simply solve the full set of Boltzmann equations.

Assuming \Eq{eq:neededcouplings} holds and plugging the relation $Y_{\psi_A} = 2 Y_{\phi_{A}}$ into the Boltzmann equations, we arrive at an effective two-particle system with $\{A,B\}$:

\begin{align}
\frac{\df Y_A}{\df x}= -\frac{\lambda}{x^2} \Biggl[ & \langle \sigma_{A \overline{A} \ra \phi \phi} v  \rangle (Y_A Y_{\Abar} - Y_{A}^{\rm{eq}} Y_{\Abar}^\text{eq}  ) \nonumber \\
\qquad \qquad ~+ \, & \langle \sigma_{\psi_A \phi_{A} \ra B \phi} v   \rangle \Biggl(Y_A^2 - Y_{B} \frac{ ( Y_A^\eq  )^2}{Y_{B}^{\rm{eq}}  } \Biggr)  \label{eq:BoltAf} \nonumber \\ 
\qquad \qquad ~  - \, & \langle \sigma_{\phi_{A}^\da B \ra \psi_A \phi} v   \rangle \Biggl(  Y_{\Abar} Y_B - Y_{A}  \frac{Y_B^\eq  Y_{\Abar}^\eq }{ Y_{A}^\eq}    \Biggr)  \nonumber \\
  \qquad \qquad ~  + 2\,& \langle \sigma_{ \psi_A \ov{B} \ra \phi_{A}^\da \phi  } v \rangle \Biggl(   Y_{A} Y_{\ov{B}}   - Y_{\Abar}    \frac{ Y_{A}^\eq Y_{\ov{B}}^\eq}{Y_{A}^\eq}       \Biggr)      \Biggr] ,  \\
  \frac{\df Y_B}{\df x} = -\frac{\lambda}{x^2} \Biggl[ &\langle \sigma_{B B^\da \ra \phi\phi} v \rangle (Y_B Y_{\Bbar} - Y_{B}^\eq Y_{\Bbar}^\eq) \nonumber \\
  \qquad \qquad ~- \, & \langle \sigma_{\psi_A \phi_{A} \ra B \phi} v \rangle \left(Y_A^2 - Y_{B} \frac{ ( Y_{A}^\eq   )^2}{Y_{B}^{\rm{eq}}  } \right)  \nonumber \\
\qquad \qquad ~ + 2 \, & \langle \sigma_{ \phi_A^\da B \ra \psi_{A} \phi  } v  \rangle  \Biggl( Y_{\ov{A}} Y_{B}   - Y_{A}  \frac{ Y_{\Abar}^\eq Y_{B}^\eq }{Y_{A}^\eq}   \Biggr) \nonumber \\   
\qquad \qquad ~+ \,&  \langle \sigma_{B \overline{q} \rightarrow qq} v \rangle  Y_\text{SM}^\eq \lp Y_B - Y_B^\eq \rp   \Biggr]    - \frac{\Gamma_B x }{H} \lp Y_B - Y_B^{\text{eq}}\rp  \label{eq:BoltBf}.
\end{align}
with conjugate equations for $\Abar$ and $\Bbar$.  Here we refer to $Y_{\psi_A}$ as simply $Y_A$ for simplicity, but we maintain old notation for the cross section in order to maintain the correct factors of $2$ in averaging over polarizations.

As one can check, the $B$ decay term and the $Bq\rightarrow qq$ scattering term depend explicitly on the chemical potentials, as expected from the discussion in \Sec{sec:chemdecouplingB}.  Below $T_D$, though, when the scattering term can be neglected, the chemical potential factors cancel in the Boltzmann system, since we take ratios of the equilibrium densities except for the decay term.  In other words, after $T_D$, we could make the replacement  $Y_{X}^\eq \to Y_{X0}^\eq$, as defined in \Eq{eq:equilibrium0}.  Numerically this is equivalent to setting the chemical potentials solely as initial conditions, as given by, say, \Eq{eq:etainitial}.

\bibliography{DarkNucleiBiB}{}
\bibliographystyle{JHEP}

\end{document}